\documentclass[%
 reprint,
 amsmath,amssymb,
pra,
]{revtex4}
\usepackage[section]{placeins}
\usepackage{graphicx}
\usepackage{dcolumn}
\usepackage{bm}

\begin{document}

\preprint{APS/123-QED}
\title{Phonon as environmental disturbance in three level system}

\author{Dawit Hiluf }
\email{dawit@post.bgu.ac.il}
\affiliation{Ben-Gurion University of Negev department of Physical Chemistry, Be'er-Sheva 84105, Israel}
\affiliation{Physics Department, Mekelle University, P.O.Box 231, Mekelle, Ethiopia.}
\author{Yonatan Dubi}
\email{jdubi@bgu.ac.il}
\affiliation{Ben-Gurion University of Negev department of Physical Chemistry, Be'er-Sheva 84105, Israel}
\affiliation{{Ilse-Katz Institute for Nanoscale Science and Technology, Ben-Gurion University of the Negev, Beer-Sheva 84105, Israel}}
\date{\today}
\begin{abstract}
This work investigates the effect of phonon coupling on the transfer of population and creation of coherence using variant of stimulated Raman adiabatic passage (STIRAP) known as \emph{fractional} stimulated Raman adiabatic passage (FSTIRAP). The study is based on the Liouville equation, which is solved numerically in the adiabatic limit. Although the phonon is assumed to be coupled only to the intermediate state, it is coupled to the other two states by dipolar system-environment interaction, inducing phonon coupling to the other states which are not directly in contact with the phonon.  At zero temperature the STIRAP pulse protocol's efficiency of the transfer decreases exponentially with the electron-phonon coupling, until the coupling strength is strong enough to make the process fully incoherent, in which case the population transfer is $\frac{1}{3}$ in each level. For the FSTIRAP protocol we find that  the transferred population to target state  decreases, leaving some population on the intermediate state. Consequently, there is an increase in the magnitude of the coherences  $\rho_{01}, \rho_{12}$, albeit small. Furthermore population transfer for non-zero temperature and effect of coupling strength is investigated, it is observed that while both parameters negatively influence the efficiency of transfer the former decrease the transfer exponentially, thereby equilibrating the system fast, while the latter seen to decrease the transfer monotonically, and hence equilibrates slowly.
\end{abstract}

\maketitle


\section{Introduction}
 Following the progress of techniques for complete population transfer, between two quantum states, via Stimulated Raman Adiabatic Passage (STIRAP) protocol, it is proposed and experimentally verified \cite{Vitanov:2001aa,Beil:2009aa}, that coherence between hyperfine levels can be created by \emph{fractional}-Stimulated Raman adiabatic passage (FSTIRAP) which is derivative of STIRAP  -- an adiabatic process which provides complete coherent transfer in a $\Lambda$-type quantum system. For an isolated system which initially is prepared to be on the ground state $|0\rangle$, one has the ability to transfer 50\% of the population to state $|2\rangle$ using  FSTIRAP, consequently creating coherence between states $|0\rangle\leftrightarrow|2\rangle$. Equipped with these two pulse protocols, i.e FSTIRAP and STIRAP, one has the ability to manipulate population transfer thereby realizing  creation and transfer of coherence. 
 
The system can equally well is considered to be a triple quantum dot (TQD) array, sharing a single electron \cite{klein2007transcending,Greentree:2004aa}. The physical system will be a linear array of three dots with electrical, rapid population transfer between any two adjacent dots. In which case one would explore electrical, rapid population transfer between two adjacent dots, such protocol is termed as coherent tunneling adiabatic passage (CTAP).  In the bra-ket notation  a state of the system is the variable $|m\rangle\langle n|$ where the TQD indices $m$ and $n$  can assume the values $0$, $1$, or ~$2$ . If $m=n$  the variable is the charge on the $m^{th}$ quantum dot (QD) and if $m\neq n$ it is the bond order between two QD. Here it has to be noted that the use of bond orders as observables follows Mulliken \cite{mulliken1955electronic,CPHC:CPHC201700222} who introduced the notion of population analysis using the charge and bond order matrix. It is argued that the quantities identified by Mulliken are the observables, populations and coherences, discussed in this paper. It should thereby be clear that by monitoring bond orders the number of state variables increases from 3 to 9. For an isolated device the total charge is conserved it then follows that, considering normalization, there are only 2 charge variables and the number of observables raises from 2 to 8 state variables. A side gate is applied whose effect on the dopant decreases with distance. Since the dopants are not equidistant from one another their coupling is not the same but depends on their separation. Such system yields electrical analogue of STIRAP and as an application of such physical system Klien et. al. designed ternary logic gates \cite{klein2007transcending}.
 
 The present authors showed the transfer of populations and coherences serving as memory in logic machine \cite{DahJoni17}. The protocol requires coupling the states, by applying laser pulses, counter intuitively, this is to say that first couple the empty states $|1\rangle$ and $|2\rangle$ using the Stokes pulse, and later drive states  $|0\rangle$ $\& ~|1\rangle$, using pump laser. If, however, the system is open to interaction with the environment the efficiency of the transfer is affected negatively.  A three level system coupled to a phonon is considered to employ the Davies-Sphon formalism in the adiabatic limits, the formalism yields an exact master equation for a weakly coupled environment \cite{breuer2007theory}.

Recently coherence is proposed for use in quantum information \cite{MichaelChuang2010} and classical logic in design of nano-scale finite state machines\cite{Fresch:2013aa}.  It then naturally follows that understanding coherence and thereby able to minimize if not avoid decoherence resulting due to either collisions or relaxations becomes a much sought question in the quantum computing realm. As a result of which there is ongoing effort to control and thereby elongate the lifetime of the coherence. To this end it therefore is logical and necessary to asses if such coherences survives the inclusion of noise. Moreover the paper will try to address the question: what impact will a phonon reveal if selectively coupled to specific atomic level.   

In this paper, we examine the effect of a phonon coupling on the dynamics of a three-level $\Lambda$-system. In addition to the phonon coupling the system is driven by a two laser fields which couples to both transitions of the $\Lambda$- system. Using the master-equation of the Davies-Sphon formalism, the populations of the each levels and the coherences created between the levels are numerically calculated. It is shown that the phonon induces coupling to states which are not directly in contact to the phonon.

The paper is organized as follows. In Sec. (\ref{sec:Hamiltonian}), the system is introduced and the Hamiltonian is derived. Next the Hamiltonian is used to obtain the master equation -- with the Davies-Sphon formalism, in the adiabatic limit, describing the evolution of the density matrix in Sec. (\ref{sec:applyDaviesSphon}) with equation of motion in the adiabatic basis given in Sec. (\ref{sec:EquationofMotion}).  Results and discussion are presented in Sec. (\ref{sec:resultsdiscussion}) before finally summarizing in Sec. (\ref{sec:conclusion}).
\section{The system and its Hamiltonian}
\label{sec:Hamiltonian}
\begin{figure}[htbp]
\begin{center}
\includegraphics[width=3.0 in]{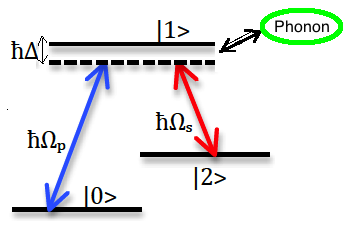}
\caption{Three level $\Lambda$ scheme coupled to a phonon, with coupling lasers. Where blue ($\Omega_P$) pump laser, and red ($\Omega_S$) stokes laser.  Detuning $\hbar\Delta_{p/s}=\hbar\omega_{jk}-\hbar\omega_{p/s}$ where $jk=01, 12$. The case $\Delta_{p}=\Delta_{s}=\Delta$ is considered.}  
\label{3levelphonon}
\end{center}
\end{figure}

The system considered is a three level system shown in Fig.(\ref{3levelphonon}) in a $\Lambda$ scheme quantum system of states $|0\rangle$, $|1\rangle$, $\& ~|2\rangle$ with energies $\hbar \omega_0$, $\hbar \omega_1$,  and $\hbar \omega_2$  respectively. These three levels  are coupled by two lasers. A pump pulse drives the transition between states $|0\rangle$ and $|1\rangle$. A second pulse, the Stokes pulse, drives the transition between $|1\rangle$ and $|2\rangle$.   Therefore, transitions between states $|0\rangle$ and $|1\rangle$, and between states $|1\rangle$ and $|2\rangle$ are allowed, but  no transition between states $|0\rangle$ $\& ~|2\rangle$, it is dipole forbidden.  \cite{Vitanov:2001aa,Shore:2008aa}. Moreover it is assumed here that a phonon is solely coupled to the intermediate state $|1\rangle$
\graphicspath{{Figures//}}

 For quantum systems with distinct states, external perturbations change the state of the system.  The three states of the system, i.e. $|0\rangle$, $|1\rangle$, and $~|2\rangle$ respectively are the eigenstates of the the unperturbed part of the Hamiltonian $\hat H_0$, and $\hat H_I$ is the part of the Hamiltonian representing the interaction of the system with the laser field. Therefore, the total Hamiltonian can be expressed as the sum of these different parts \cite{Shore:2008aa}.
\begin{equation}
 \begin{aligned}
\hat H= &\hat H_{sys}+\hat H_{int}+\hat H_{ph}+\hat H_{sys-ph}
\end{aligned}
\end{equation}
where $\hat H_{sys},H_{int}, \hat H_{ph}, \hat H_{sys-ph}$ are, respectively, system Hamiltonian, interaction Hamiltonian between the system and the laser pulse, free Hamiltonian of the phonon,  and interaction between system and phonon part of the Hamiltonian and are expressed as 
\begin{subequations}
\begin{align}
\hat H_{sys}=&\sum_{n=0}^2\hbar\omega_n|n\rangle\langle n|\label{Hsys}\\
\hat H_{int}=&\Omega_P\left(|0\rangle\langle 1|+|1\rangle\langle 0|\right)+\Omega_S\left(|1\rangle\langle 2|+|2\rangle\langle 1|\right)\label{Hint}\\
\hat H_{ph}=&\omega_{ph}\hat a^\dagger\hat a\label{Hph}\\ 
\hat H_{sys-ph}=&\lambda\left(\hat a^\dagger+\hat a\right)|1\rangle\langle 1|\label{Hsysph}
\end{align}
\end{subequations}
where $\hbar\omega_n$ is energy of level $n=0, 1, 2$, $\Omega_k, k=P,S$ is the laser pulse driving the transitions $|0\rangle\leftrightarrow|1\rangle,~|1\rangle\leftrightarrow|2\rangle$, $\omega_{ph}$ is frequency of the phonon, $\hat a^\dagger, \hat a$ are the creation and annihilation operators of the phonon, and $\lambda$ is the coupling strength of the phonon to level $|1\rangle$. At this point the system-phonon must be decoupled. This can be done by following Lang and Frisov recommendation and make the transformation $\tilde H=e^S H e^{-S}$. Obviously Hermiticity of the Hamiltonian has to be preserved, and for this to happen the transformation must be unitary. This consequently implies that the generators $S$ have to be anti-Hermitian, i.e $S^\dagger=-S$, Eq.\eqref{polaron} gives appropriate choice of  $S$ that satisfies these requirements  
\begin{equation}
\begin{aligned}
S= \frac{\lambda}{\omega_{ph}}\left(\hat a^\dagger-\hat a\right)|1\rangle\langle 1|
\end{aligned}
\label{polaron}
\end{equation}
Next the Baker-Campbell-Hausdorff formula is applied to transform Eqs.\eqref{Hsys},\eqref{Hint},\eqref{Hph}, and\eqref{Hsysph},  into
\begin{equation}
\begin{aligned}
\tilde H=&\tilde H_{sys}+\tilde H_{B}+\tilde V
 \end{aligned}
\end{equation}
with $\tilde V$ the interaction part of the Hamiltonian and where
\begin{subequations}
\begin{align}
\tilde H_{sys}=&\sum_{n=0}^2E_n|n\rangle\langle n|-\frac{\lambda^2}{\omega_{ph}}|1\rangle\langle 1|\label{Eq:reorg}\\
\tilde H_{B}=&\omega_{ph}\hat a^\dagger\hat a\\
\begin{split}
\tilde V=&\Omega_P\Big(e^{\frac{\lambda}{\omega_{ph}}\left(\hat a^\dagger-\hat a\right)}|1\rangle\langle 0|+e^{-\frac{\lambda}{\omega_{ph}}\left(\hat a^\dagger-\hat a\right)}|0\rangle\langle 1|\Big)\\
&+\Omega_S\Big(e^{\frac{\lambda}{\omega_{ph}}\left(\hat a^\dagger-\hat a\right)}|1\rangle\langle 2|+e^{-\frac{\lambda}{\omega_{ph}}\left(\hat a^\dagger-\hat a\right)}|2\rangle\langle 1|\Big)\label{Eq:vtilde}
\end{split}
 \end{align}
\end{subequations}
Note here that Eq.\eqref{Eq:reorg} is the reorganisation energy shift and Eq.\eqref{Eq:vtilde} informs one that the state $|1\rangle$ is coupled to the other two states by phonon induced system-environment interaction. Recall it was assumed the phonon coupling to be solely with state $|1\rangle$, but interestingly, as seen in  Eq.\eqref{Eq:vtilde}, the vibrating level due to the coupling phonon also affects the other two states, i.e. $|0\rangle, |2\rangle$. 

 Equation \eqref{Eq:vtilde} needs further decoupling, as it embodies phonon induced-system coupling, by employing the Born approximation. In this context the Born approximation corresponds to first order term in expansion of Eq.\eqref{Eq:vtilde} in $\lambda$, meaning weak interaction between the system and the phonon reservoir is assumed so that one-phonon transition is dominant in comparison to multi-phonon transition. expanding $\tilde V$ up to first order leads to the following expression for the interaction part
 \begin{widetext}
\begin{equation}
\begin{aligned}
\tilde V=&\hbar\Omega_P\left(t\right)\Big[\left(1+\frac{\lambda}{\omega_{ph}}\left(\hat a^\dagger-\hat a\right)\right)|1\rangle\langle 0|+\left(1-\frac{\lambda}{\omega_{ph}}\left(\hat a^\dagger-\hat a\right)\right)|0\rangle\langle 1|\Big]\\
+&\hbar\Omega_S\left(t\right)\Big[\left(1+\frac{\lambda}{\omega_0}\left(\hat a^\dagger-\hat a\right)\right)|1\rangle\langle 2|+\left(1-\frac{\lambda}{\omega_{ph}}\left(\hat a^\dagger-\hat a\right)\right)|2\rangle\langle 1|\Big]
 \end{aligned}
\end{equation}
which then yields 
\begin{equation}
\begin{aligned}
\tilde V=&\hbar\Omega_P\left(t\right)\left(|1\rangle\langle 0|+|0\rangle\langle 1|\right)+\frac{\lambda\left(t\right)}{\omega_{ph}}\left(\hat a^\dagger-\hat a\right)\left(|1\rangle\langle 0|-|0\rangle\langle 1|\right)\\
+&\hbar\Omega_S\left(t\right)\left(|1\rangle\langle 2|+|2\rangle\langle 1|\right) +\frac{\lambda\left(t\right)}{\omega_{ph}}\left(\hat a^\dagger-\hat a\right)\left(|1\rangle\langle 2|-|2\rangle\langle 1|\right) 
\end{aligned}
\label{Eq:Vtildeeq}
\end{equation}
\end{widetext}
where $\lambda\left(t\right)=\lambda\left(\hbar\Omega_k\left(t\right)\right)$ with $k=P,S$. It is worth pointing out here that the pulses have phase, i.e. $\hbar\Omega_k\left(t\right)=\hbar\Omega_k \left(t\right)e^{\pm i\omega_kt}$, where $\omega_k, k=P,S$ is the frequency of pump laser and Stokes laser respectively and is given by 
\begin{equation}
\begin{aligned}
\omega_P=\omega_1-\omega_0-\Delta_P=\omega_{10}-\Delta\\
\omega_S=\omega_1-\omega_2-\Delta_S=\omega_{12}-\Delta\\
\end{aligned}
\label{omegaPS}
\end{equation}
where $\omega_{mn}=\omega_m-\omega_n$ and  single photon detuning $\Delta_P=\Delta_S=\Delta$ is assumed. Thus considering the phases of the laser pulses, and Eq.\eqref{omegaPS}, one can rewrite Eq.\eqref{Eq:Vtildeeq} as  
\begin{widetext}
\begin{equation}
\begin{aligned}
\tilde V=&\hbar\Omega_P\left(e^{i(\omega_{10}-\Delta)t}|0\rangle\langle 1|+e^{ -i(\omega_{10}-\Delta)t}|1\rangle\langle 0|\right)+\hbar\Omega_S\left(e^{- i(\omega_{12}-\Delta)t}|1\rangle\langle 2|+e^{ i(\omega_{12}-\Delta)t}|2\rangle\langle 1|\right)\\
+& \frac{\lambda\left(t\right)}{\omega_{ph}}\left(\hat a^\dagger-\hat a\right)\left(|1\rangle\langle 0|-|0\rangle\langle 1|\right)+\frac{\lambda\left(t\right)}{\omega_{ph}}\left(\hat a^\dagger-\hat a\right)\left(|1\rangle\langle 2|-|2\rangle\langle 1|\right) 
\end{aligned}
\label{Eq:Vtildeeq3}
\end{equation}
\end{widetext}
In the rotating frame associated with the transformation $e^{i\omega_{0}t}|0\rangle\langle 0|+e^{-i(\omega_{1}-\Delta)t}|1\rangle\langle 1|+e^{i\omega_{2}t}|2\rangle\langle 2|$\cite{Scala2011}, the total Hamiltonian takes the form
\begin{equation}
\begin{aligned}
H\left(t\right)=&H_{sys}\left(t\right)+H_B+V\left(t\right)
\end{aligned}
\end{equation}
with 
\begin{equation}
\begin{aligned}
H_{sys}=&\hbar\tilde\Delta|1\rangle\langle 1|+\hbar\Omega_P\left(t\right)\left(|0\rangle\langle 1|+|1\rangle\langle 0|\right)\\ &+\hbar\Omega_S\left(t\right)\left(|1\rangle\langle 2|+|2\rangle\langle 1|\right)\\
H_{B}=&\omega_{ph}\hat a^\dagger\hat a\\
V\left(t\right)=& \frac{\lambda\left(t\right)}{\omega_{ph}}\left(\hat a^\dagger-\hat a\right)\left(e^{-i(\omega_{01}+\Delta)t}|1\rangle\langle 0|-e^{i(\omega_{01}+\Delta)t}|0\rangle\langle 1|\right)\\
+&\frac{\lambda\left(t\right)}{\omega_{ph}}\left(\hat a^\dagger-\hat a\right)\left(e^{-i(\omega_{21}+\Delta)t}|1\rangle\langle 2|-e^{i(\omega_{21}+\Delta)t}|2\rangle\langle 1|\right)
\end{aligned}
\label{Eq:TotHamilRWA}
\end{equation}
where $\omega_{jk}=\omega_j-\omega_k$ and $\hbar\tilde\Delta=\left(\hbar\Delta-\frac{\lambda^2}{\omega_{ph}}\right)$ (for small $\lambda$ it follows $\lambda\rightarrow0$ and thus $\tilde\Delta=\Delta$).
Thus with Eq.\eqref{Eq:TotHamilRWA} it is seen that the system and the environment Hamiltonian have been decoupled. After diagonalising the system Hamiltonian one readily finds the eigenvalues  $\{\omega_+,\omega_0,\omega_-\}$ to be
\begin{subequations}. 
\begin{align}
\omega_+=&\frac{1}{2}\big(\Delta+\sqrt{\Delta^2+\Omega^2}\big)=\Omega\cot\phi,\\
\omega_0=&0, \\
\omega_-=&\frac{1}{2}\big(\Delta-\sqrt{\Delta^2+\Omega^2}\big)=-\Omega\tan\phi
\label{eigvalue}
\end{align}
\end{subequations} 
with $\Omega^2=\Omega_P^2+\Omega_S^2$
where the mixing angles are defined to be 
\begin{subequations}
\begin{align}
\tan\theta=&\frac{\Omega_P\left(t\right)}{\Omega_S\left(t\right)},\label{angle1a}\\
\tan 2\phi=&\frac{\Omega\left(t\right)}{\Delta}\label{angle2a}
\end{align}
\end{subequations} 
Diagonalising the system Hamiltonian also yields new basis vectors corresponding to the eigenvalues $\{\omega_+,\omega_0,\omega_-\}$ respectively, i.e $\{|a_+\rangle,|a_0\rangle,|a_-\rangle\}$,  where
\begin{subequations}
\begin{align}
|a_+\rangle=&\sin\theta\sin\phi |0\rangle+\cos\phi  |1\rangle+ \cos\theta\sin\phi |2\rangle \label{estate1}\\
|a_0\rangle=&\cos\theta |0\rangle-\sin\theta  |2\rangle \label{estate2}\\
|a_-\rangle=&\sin\theta\cos\phi |0\rangle-\sin\phi  |1\rangle+ \cos\theta\cos\phi |2\rangle \label{estate3}
\end{align}
\end{subequations}
Notice these new basis reveals interesting picture as in this new picture the adiabatic state $|a_0\rangle$ is independent of state $|1\rangle$, recall also that state $|1\rangle$ is the state directly in contact with the phonon and it is the leaky state. Therefore following the adiabatic state $|a_0\rangle$ the complete population transfer  from the ground state $|0\rangle$ to the target state $|2\rangle$ can be achieved. 
 Note also that the system Hamiltonian is varying slowly whereas the system-environment is  time-dependent and is oscillating. One way of addressing such system is to employ the Davies-Sphon formalism which deals such system-environment Hamiltonian \cite{breuer2007theory}, the next section makes use of the equation of motion derived using the approach and Appendix (\ref{sec:applyDaviesSphon}) provides brief recap of the formalism.

\section{Equation of motion in the adiabatic basis}
\label{sec:EquationofMotion}
It is worth pointing out that in the rotating framework with a slowly-varying system Hamiltonian and a time dependent system-environment interaction term consequently one can employ the general formalism of Davies and Sphon to derive the master equation. To this end regard the system Hamiltonian $H_{sys}$ as constant and the system is interacting with an environment which contains an oscillating term. Because the environmental correlation time is much smaller than the timescales of the Hamiltonian change, it can therefore be treated as if the system Hamiltonian is constant by assuming that the system Hamiltonian is time-independent during the derivation of the master equation and including the time dependence of the transition operators latter.  Later in the next section we will employ these formalism, i.e Davies-Sphon to the system at hand.

Therefore upon using Eq.\eqref{rhodot5} along with the transition operators obtained in Appendix (\ref{sec:jumpingOperators}), it is readily obtain that
\begin{widetext}
\begin{equation}
\begin{aligned}
\dot\rho=&-i\left[H_s,\rho\right]\\
+&\left(\gamma_{aa}^{++}\left(\omega_{+0}\right)\cos^2\theta\cos^2\phi+\gamma_{bb}^{++}\left(\omega_{+0}\right)\sin^2\theta\cos^2\phi\right)\left(|a_0\rangle\langle a_+|\rho|a_+\rangle\langle a_0|-\frac{1}{2}\{|a_+\rangle\langle a_+|,\rho\}\right)\\
+&\left(\gamma_{aa}^{--}\left(\omega_{0-}\right)\cos^2\theta\sin^2\phi+\gamma_{bb}^{--}\left(\omega_{0-}\right)\sin^2\theta\sin^2\phi\right)\left(|a_-\rangle\langle a_0|\rho|a_0\rangle\langle a_-|-\frac{1}{2}\{|a_0\rangle\langle a_0|,\rho\}\right)\\
+&\left(\gamma_{aa}^{++}\left(\omega_{+-}\right)\sin^2\theta\cos^4\phi+\gamma_{aa}^{--}\left(\omega_{+-}\right)\sin^2\theta\sin^4\phi\right)\left(|a_-\rangle\langle a_+|\rho|a_+\rangle\langle a_-|-\frac{1}{2}\{|a_+\rangle\langle a_+|,\rho\}\right)\\
+&\left(\gamma_{bb}^{++}\left(\omega_{+-}\right)\cos^2\theta\cos^4\phi+\gamma_{bb}^{--}\left(\omega_{+-}\right)\cos^2\theta\sin^4\phi\right)\left(|a_-\rangle\langle a_+|\rho|a_+\rangle\langle a_-|-\frac{1}{2}\{|a_+\rangle\langle a_+|,\rho\}\right)\\
+&\left(\gamma_{aa}^{++}\left(0\right)+\gamma_{aa}^{--}\left(0\right)\right)\sin^2\theta\sin^2\phi\cos^2\phi\\
 \times&\left(\left(|a_+\rangle\langle a_+|-|a_-\rangle\langle a_-|\right)\rho\left(|a_+\rangle\langle a_+|-|a_-\rangle\langle a_-|\right)-\frac{1}{2}\{|a_+\rangle\langle a_+|+|a_-\rangle\langle a_-|,\rho\}\right)\\
 +&\left(\gamma_{bb}^{++}\left(0\right)+\gamma_{bb}^{--}\left(0\right)\right)\cos^2\theta\sin^2\phi\cos^2\phi\\
 \times&\left(\left(|a_+\rangle\langle a_+|-|a_-\rangle\langle a_-|\right)\rho\left(|a_+\rangle\langle a_+|-|a_-\rangle\langle a_-|\right)-\frac{1}{2}\{|a_+\rangle\langle a_+|+|a_-\rangle\langle a_-|,\rho\}\right)\\
+&\left(\gamma_{aa}^{--}\left(\omega_{0+}\right)\cos^2\theta\cos^2\phi+\gamma_{bb}^{--}\left(\omega_{0+}\right)\sin^2\theta\cos^2\phi\right)\left(|a_+\rangle\langle a_0|\rho|a_0\rangle\langle a_+|-\frac{1}{2}\{|a_0\rangle\langle a_0|,\rho\}\right)\\
+&\left(\gamma_{aa}^{++}\left(\omega_{-0}\right)\cos^2\theta\sin^2\phi+\gamma_{bb}^{++}\left(\omega_{-0}\right)\sin^2\theta\sin^2\phi\right)\left(|a_0\rangle\langle a_-|\rho|a_-\rangle\langle a_0|-\frac{1}{2}\{|a_-\rangle\langle a_-|,\rho\}\right)\\
+&\left(\gamma_{aa}^{--}\left(\omega_{-+}\right)\sin^2\theta\cos^4\phi+\gamma_{aa}^{++}\left(\omega_{-+}\right)\sin^2\theta\sin^4\phi\right)\left(|a_+\rangle\langle a_-|\rho|a_-\rangle\langle a_+|-\frac{1}{2}\{|a_-\rangle\langle a_-|,\rho\}\right)\\
+&\left(\gamma_{bb}^{--}\left(\omega_{-+}\right)\cos^2\theta\cos^4\phi+\gamma_{bb}^{++}\left(\omega_{-+}\right)\cos^2\theta\sin^4\phi\right)\left(|a_+\rangle\langle a_-|\rho|a_-\rangle\langle a_+|-\frac{1}{2}\{|a_-\rangle\langle a_-|,\rho\}\right)\\
\end{aligned}
\label{rhodotadia}
\end{equation}
\end{widetext}
where $\omega_{jk}=\omega_j-\omega_k$
From Eqs.\eqref{Gammapp} and \eqref{Gammamm} one gets that the decay rates are given by a spectral density $J_j\left(\omega\right)$ multiplied by factor depending on the photon population $N\left(\omega\right)$ of the bath modes at the relevance frequency along with $\pm\omega_a$ or $\pm\omega_b$ used accordingly
\begin{equation}
\begin{aligned}
\gamma_{jj}^{++}\left(\omega\right)=&J_j\left(\omega-\omega_j\right)\left(1+N\left(\omega-\omega_j\right)\right)  && \omega-\omega_j > 0\\
\gamma_{jj}^{--}\left(\omega\right)=&J_j\left(|\omega+\omega_j|\right)\left(N\left(|\omega+\omega_j|\right)\right)  && \omega+\omega_j < 0
\end{aligned}
\label{meanN}
\end{equation}
with $j=a, b$ and
\begin{equation}
\begin{aligned}
\omega_a=&\omega_0-\omega_1+\Delta\\
\omega_b=&\omega_2-\omega_1+\Delta
\end{aligned}
\label{omwgeaw}
\end{equation}
The measure of system-environment coupling strength at zero temperature, i.e. spectral density $J\left(\omega\right)$ for general reservoir is given by
\begin{equation}
\begin{aligned}
J\left(\omega\right)=&\sum_k\Big|\frac{\lambda}{\omega_{ph}}\Big|^2\delta(\omega-\omega_k)
\end{aligned}
\label{OmegaJw}
\end{equation}
where $\lambda$ is system-environment coupling. 

It is worth pointing out that the time integral in Eqs.\eqref{Gammapp} and \eqref{Gammamm} cab be done using the relation 
\begin{equation}
\begin{aligned}
\int_0^\infty ds e^{-ixs}=&\pi\delta(x)-iP\left(\frac{1}{x}\right)
\end{aligned}
\label{InteRela}
\end{equation}
here $P$ denotes the Cauchy principal value. From which follows that the real part of the integral gives rise to environment-induced decay of the system whereas the imaginary parts cause environment-induced energy shift. Moreover  the decay rate is directly proportional to the spectral density evaluated at the system energy splitting. This in turn depends on the details of the environment coupling parameters. If one assumes flat spectrum it implies that $J_j=\Gamma$, where $\Gamma$ is life time of the excited state. In this paper the environment induced energy shift commonly referred to as Lamb-shift arising from the imaginary part of integral is neglected. Furthermore in the numerical calculations it is assumed $\Gamma_{01}=\Gamma_{21}=\Gamma$.


 At zero temperature the number of photons in the reservoir is zero and thus $\gamma_{aa}^{--}=\gamma_{bb}^{--}=0$. In doing so it is revealed that there is transitions from $|a_+\rangle$ to $|a_-\rangle$ and vice versa, transitions from $|a_+\rangle ~\&~ |a_-\rangle$ to $|a_0\rangle$. This implies that there is dephasing process involving levels $|a_+\rangle$ and $|a_-\rangle$. In a nutshell all this suggests that damping can help to transfer population to level $|a_0\rangle$, so that the efficiency of population transfer be affected positively.
In this section the explicit form of the equation of the system is provide.  The time evolution of the density matrix in time in Hilbert space represents a system of $N^2$ coupled linear differential equations for the individual density matrix elements. It is worth pointing out that the $N\times N$ elements of the density matrix in Hilbert space can be arranged in Liouville space vector of length $N^2$. In doing so, the equation of motions can now be expressed in dimension  of $N^2\times N^2$  in Liouville  space. To this end  Eq.\eqref{rhodotadia} takes the following form
\begin{widetext}
 \begin{equation}
\begin{aligned}
\dot\rho=&-i\left[H_s,\rho\right]\\
+&\gamma_1\left(|a_0\rangle\langle a_+|\rho|a_+\rangle\langle a_0|-\frac{1}{2}\{|a_+\rangle\langle a_+|,\rho\}\right)
+\gamma_2\left(|a_0\rangle\langle a_-|\rho|a_-\rangle\langle a_0|-\frac{1}{2}\{|a_-\rangle\langle a_-|,\rho\}\right)\\
+&\gamma_3\left(|a_-\rangle\langle a_+|\rho|a_+\rangle\langle a_-|-\frac{1}{2}\{|a_+\rangle\langle a_+|,\rho\}\right)
+\gamma_4\left(|a_+\rangle\langle a_-|\rho|a_-\rangle\langle a_+|-\frac{1}{2}\{|a_-\rangle\langle a_-|,\rho\}\right)\\
+&\gamma_5\left(\left(|a_+\rangle\langle a_+|-|a_-\rangle\langle a_-|\right)\rho\left(|a_+\rangle\langle a_+|-|a_-\rangle\langle a_-|\right)-\frac{1}{2}\{|a_+\rangle\langle a_+|+|a_-\rangle\langle a_-|,\rho\}\right)\\
+&\gamma_{6}\left(|a_-\rangle\langle a_0|\rho|a_0\rangle\langle a_-|-\frac{1}{2}\{|a_0\rangle\langle a_0|,\rho\}\right)\\
+&\gamma_{7}\left(|a_+\rangle\langle a_0|\rho|a_0\rangle\langle a_+|-\frac{1}{2}\{|a_0\rangle\langle a_0|,\rho\}\right)
 \end{aligned}
\label{rhodotadiazero1}
\end{equation}
\end{widetext}
where 
\begin{widetext}
\begin{equation}
\begin{aligned}
\gamma_1=&\left(\gamma_{aa}^{++}\left(\omega_{+0}\right)\cos^2\theta\cos^2\phi+\gamma_{bb}^{++}\left(\omega_{+0}\right)\sin^2\theta\cos^2\phi\right)\\
\gamma_2=&\left(\gamma_{aa}^{++}\left(\omega_{-0}\right)\cos^2\theta\sin^2\phi+\gamma_{bb}^{++}\left(\omega_{-0}\right)\sin^2\theta\sin^2\phi\right)\\
\gamma_3=&\left(\gamma_{aa}^{++}\left(\omega_{+-}\right)\sin^2\theta\cos^4\phi+\gamma_{bb}^{++}\left(\omega_{+-}\right)\cos^2\theta\cos^4\phi\right)\\
+&\left(\gamma_{aa}^{--}\left(\omega_{+-}\right)\sin^2\theta\sin^4\phi+\gamma_{bb}^{--}\left(\omega_{+-}\right)\cos^2\theta\sin^4\phi\right)\\
\gamma_4=&\left(\gamma_{aa}^{++}\left(\omega_{-+}\right)\sin^2\theta\cos^4\phi+\gamma_{bb}^{++}\left(\omega_{-+}\right)\cos^2\theta\sin^4\phi\right)\\
+&\left(\gamma_{aa}^{--}\left(\omega_{-+}\right)\sin^2\theta\cos^4\phi+\gamma_{bb}^{--}\left(\omega_{-+}\right)\cos^2\theta\cos^4\phi\right)\\
\gamma_5=&\left(\gamma_{aa}^{++}\left(0\right)\sin^2\theta\sin^2\phi\cos^2\phi+\gamma_{bb}^{++}\left(0\right)\cos^2\theta\sin^2\phi\cos^2\phi\right)\\
+&\left(\gamma_{aa}^{--}\left(0\right)\sin^2\theta\sin^2\phi\cos^2\phi+\gamma_{bb}^{--}\left(0\right)\cos^2\theta\sin^2\phi\cos^2\phi\right)\\
\gamma_{6}=&\left(\gamma_{aa}^{--}\left(\omega_{0-}\right)\cos^2\theta\sin^2\phi+\gamma_{bb}^{--}\left(\omega_{0-}\right)\sin^2\theta\sin^2\phi\right)\\
\gamma_{7}=&\left(\gamma_{aa}^{--}\left(\omega_{0+}\right)\cos^2\theta\cos^2\phi+\gamma_{bb}^{--}\left(\omega_{0+}\right)\sin^2\theta\cos^2\phi\right)\\
\end{aligned}
\label{regamma}
\end{equation}
\end{widetext}
The density matrix can now be decomposed in terms of the instantaneous eigenstates of the Hamiltonian $H_s(t)$ as 
\begin{equation}
\begin{aligned}
\rho\left(t\right)=&\sum_{jk}\rho_{jk}\left(t\right)|a_j\left(t\right)\rangle\langle a_k\left(t\right)|
\end{aligned}
\label{rhoexpandp0m}
\end{equation}
where $j,k=+,0,-$. Using this decomposition and the Hamiltonian given in Eq.\eqref{HnonA} along with the relevant corresponding contributing terms  (of $-i\left[H_s,\rho\right]$ of Eq.\eqref{rhodotadia} or Eq.\eqref{rhodotadiazero1}), the equation of motion for the adiabatic density matrix in the ordering $\rho^a=(\rho_{+0}, \rho_{+-}, \rho_{0-}, \rho_{0+}, \rho_{-+}, \rho_{-0}, \rho_{++}, \rho_{00}, \rho_{--})^T$ takes the form
\begin{equation}
\begin{aligned}\frac{d}{dt}\rho^a=&M \rho^a
\end{aligned}
\end{equation}
where $M$ is given
\begin{widetext}
\begin{equation}
\begin{aligned}M=&
\begin{pmatrix}
-i\omega_{+0}+\Gamma_{+0} & -\dot\theta\cos\phi & 0 & 0 & 0 & \dot\phi & -\dot\theta\sin\phi & \dot\theta\sin\phi & 0\\
\dot\theta\cos\phi & -i\omega_{+-}+\Gamma_{+-} & \dot\theta\sin\phi & 0 &0 & 0 &  -\dot\phi & 0 &  \dot\phi \\
0 & -\dot\theta\sin\phi & -i\omega_{0-}+\Gamma_{0-} & -\dot\phi & 0 & 0 & 0 & \dot\theta\cos\phi & -\dot\theta\cos\phi\\
0 & 0 & \dot\phi  & i\omega_{0+}+\Gamma_{0+} & -\dot\theta\cos\phi  & 0 & -\dot\theta\sin\phi & \dot\theta\sin\phi & 0\\
0 & 0 & 0 & \dot\theta\cos\phi & i\omega_{+-}+\Gamma_{+-}  & \dot\theta\sin\phi  & -\dot\phi & 0 &  \dot\phi \\
-\dot\phi & 0 & 0 & 0 & \dot\theta\sin\phi &i\omega_{0-}+\Gamma_{-0} & 0 & \dot\theta\cos\phi & -\dot\theta\cos\phi\\
\dot\theta\sin\phi & \dot\phi  & 0 & \dot\theta\sin\phi & \dot\phi & 0 & \Gamma_{++} & \gamma_{7} & \gamma_4\\
-\dot\theta\sin\phi & 0 & -\dot\theta\cos\phi & -\dot\theta\sin\phi & 0 & -\dot\theta\cos\phi  & \gamma_1 & \Gamma_{00} & \gamma_2\\
0 & -\dot\phi & \dot\theta\cos\phi & 0 & -\dot\phi & \dot\theta\cos\phi  & \gamma_3 & \gamma_{6} & -\Gamma_{--}
\end{pmatrix}
\end{aligned}
\label{rhodotsp0m}
\end{equation}
\end{widetext}
where $\omega_{+0}=\Omega\cot\phi, \omega_{+-}=\Omega\csc\phi\sec\phi, \omega_{0-}=\Omega\tan\phi$ and the relaxation terms are
\begin{equation}
\begin{aligned}
\Gamma_{+0}=&-i\omega_{+0}-\frac{1}{2}\left(\gamma_1+\gamma_3+\gamma_5+\gamma_{6}+\gamma_{7}\right)\\
\Gamma_{+-}=&-i\omega_{+-}-\frac{1}{2}\left(\gamma_1+\gamma_2+\gamma_3+\gamma_4-4\gamma_5\right)\\
\Gamma_{0-}=&-i\omega_{0-}-\frac{1}{2}\left(\gamma_2+\gamma_4+\gamma_5+\gamma_{6}+\gamma_{7}\right)\\
\Gamma_{++}=&-\left(\gamma_1+\gamma_3\right)\\
\Gamma_{--}=&\left(\gamma_{2}+\gamma_{4}\right)\\
\Gamma_{00}=&-\left(\gamma_{6}+\gamma_{7}\right)
\end{aligned}
\label{newgammas}
\end{equation}
with $\Gamma_{jk}=\Gamma_{kj}, j,k=+,0,-$. Where the over-dot means time derivative. Differentiating Eqs.\eqref{angle1a}  and \eqref{angle2a} with respect to time and using chain rule of differentiation provides
\begin{subequations}
\begin{align}
\dot\theta=&\frac{\dot\Omega_P\Omega_S-\dot\Omega_S\Omega_P}{\Omega_P^2+\Omega_S^2} \label{tetadot}\\
\dot\phi=&\frac{1}{2}\Big[\frac{\Delta}{\Omega}\Big(\frac{\dot\Omega_P\Omega_P+\dot\Omega_S\Omega_S}{\Delta^2+\Omega^2}\Big)\Big] \label{fidot}
\end{align}
\end{subequations}
\section{Results and Discussions}
\label{sec:resultsdiscussion}
Now with the relevant equations of motion developed we proceed to apply the formalism to two protocols, STIRAP and FSTIRAP. To begin with recall though assumption was made for the phonon coupling to be solely with state $|1\rangle$, but it is to be noted from  Eq.\eqref{Eq:vtilde} that the vibrating level due to the coupling phonon also affects the other two states, i.e. $|0\rangle, |2\rangle$. This vibrating level in turn results, as seen in Eq.\eqref{Eq:reorg}, a reorganisation energy shift.  
 Moreover when temperature is zero the efficiency of population transfer is affected positively and thus in the adiabatic basis the population remains trapped in adiabatic state $|a_0\rangle$, this is so because the population transfer path, directly or indirectly, leads to state $|a_0\rangle$.  However the condition is different when when non-zero temperature is considered. 
 
 For starter STIRAP pulse profile is used, and plot of population at each level versus time is shown in Fig.(\ref{fig:GammaPopSP}), it is seen that increasing the strength of the noise leads to incoherent process as expected. The incoherent processes arises because the coupled phonon vibrates the states until the whole system equilibrates. The shaky nature of the levels in turn creates inability of complete population transfer as is the norm for STIRAP. This difficulty can be viewed in two ways, first the vibrating levels are no more able to retain the population transferred to them because the phonon is shaking the whole system. Second the path to the target state is no more achievable as it is not stable because of the coupling.  
 
\begin{widetext}
\graphicspath{{Plotss//}}
\begin{figure}[htp]
\centering
(a)\includegraphics[width=.3\textwidth]{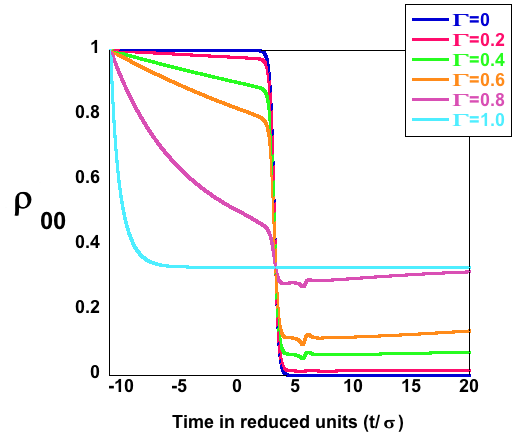}
(b)\includegraphics[width=.3\textwidth]{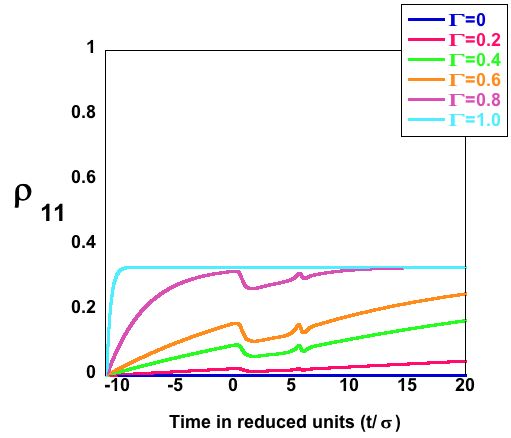}
(c)\includegraphics[width=.3\textwidth]{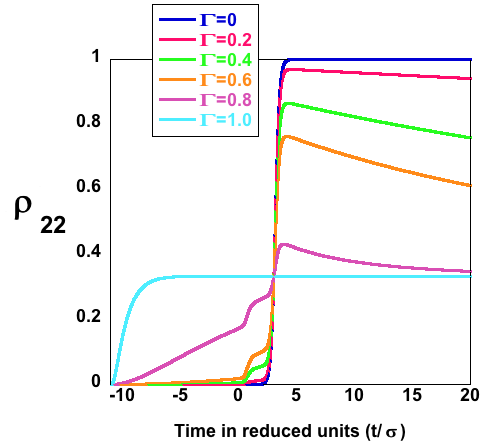}
\label{fig:GammaPopSP}
\caption{Plots of the  bare state population for different values of $\Gamma$ using STIRAP pulse protocol. System is initially prepared to be in the ground state $|0\rangle$  (a) Population in the ground state $|0\rangle$ (b) Population in the  intermediate state $|1\rangle$. (c) Population in the target state $|2\rangle$. Where the $\Gamma=0,0.2,0.4,0.6,0.8,1$ respectively represent for values $\Gamma=0, \frac{1}{500\sigma}, \frac{1}{100\sigma},\frac{1}{50\sigma},\frac{1}{10\sigma},\frac{1}{\sigma}$}
\label{fig:GammaPopSP}
\end{figure}

Next we change the protocol and we now use FSTIRAP pulse scheme. Here also plot of population of each level versus time for FSTIRAP profile is shown. In this case, i.e for FSTIRAP protocol, it is seen in Figs.~\ref{fig:GammaPopFSP} that for $\Gamma=0$ the superposition is maximum, i.e. equally distributed between the ground state $|0\rangle$ and the target state $|2\rangle$, but as the values of $\Gamma$ increase  the transferred population to target state $|2\rangle$ decreases and in doing so, along the way, there will be some population on state $|1\rangle$. 

\graphicspath{{Plotss//}}
\begin{figure}[htp]
\centering
(a)\includegraphics[width=.3\textwidth]{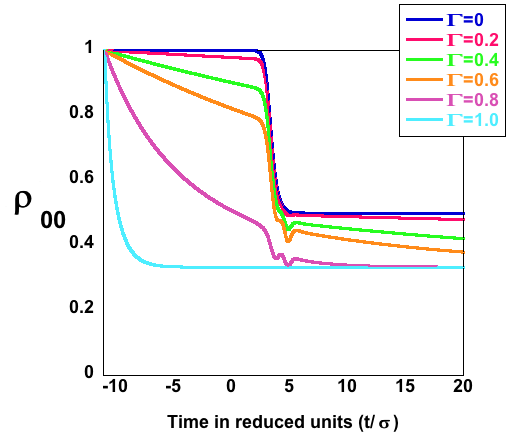}
(b)\includegraphics[width=.3\textwidth]{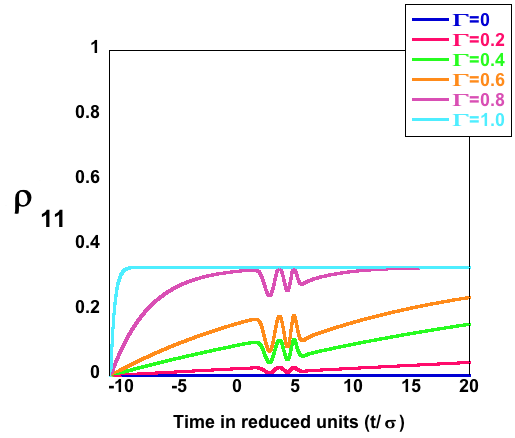}
(c)\includegraphics[width=.3\textwidth]{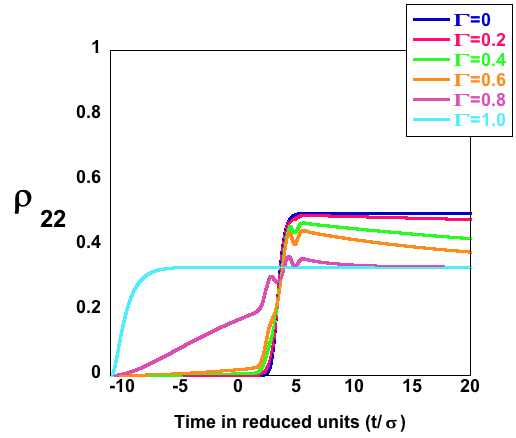}
\label{fig:GammaPopFSP}
\caption{Plots of the  bare state population for different values of $\Gamma$ using FSTIRAP pulse protocol. System is initially prepared to be in the ground state $|0\rangle$  (a) Population in the ground state $|0\rangle$ (b) Population in the  intermediate state $|1\rangle$. (c) Population in the target state $|2\rangle$. Where the $\Gamma=0,0.2,0.4,0.6,0.8,1$ respectively represent for values $\Gamma=0, \frac{1}{500\sigma}, \frac{1}{100\sigma},\frac{1}{50\sigma},\frac{1}{10\sigma},\frac{1}{\sigma}$}
\label{fig:GammaPopFSP}
\end{figure}
\end{widetext}

To explore better what is happening plots of bare state populations and coherences for different values of $\Gamma$ are investigated. It then follows that for the STIRAP pulse protocol the efficiency of the transfer decreases as the strength of the environment increase, until the strength is strong enough to  make the process incoherent in which case the population transfer is $\frac{1}{3}$ in each level.

However, the condition needs closer look up for FSTIRAP protocol, because when there is no phonon coupling the system distributes the population equally between the two ground states $|0\rangle$ and $|2\rangle$. But, as we increase the noise strength the phonon shakes the levels and manages to put some population at each level, consequently coherences [other than $\rho_{02}$] seen to be created albeit small in magnitude see Figs.(\ref{fig:GammaCohrFSP}a) and (\ref{fig:GammaCohrFSP}b). On top of this as the strength of the noise increases further, the coupling with phonon completely destroys the phase correlation, and  the coherences deteriorates. This is a clear indication and an interesting observation where moderate noise is seen creating coherences. The optimum value that enables the creation before it starts destroying the coherences is related to the time scale of the leaky level, i.e $|1\rangle$, to which the phonon is directly coupled. The effect of having population on state $|1\rangle$ consequently means that  coherences are created (consult Eq.\eqref{cohrjk}).  This phenomenon of creation of coherences can be clarified by loosely defining, ignoring the phases, the coherences as
\begin{equation} 
\begin{aligned}
|\rho_{jk}|=|\rho_{jj}||\rho_{kk}|
\end{aligned}
\label{cohrjk}
\end{equation}
Therefore it follows from Eq.\eqref{cohrjk} that if one manages to put measurable amount of populations in both states $|j\rangle, |k\rangle$ thereby yielding population at each state to be $\rho_{jj} ~\&~ \rho_{kk}$, the existence of population at each level consequently results in a coherence term between the two levels $|j\rangle, |k\rangle$, i.e $\rho_{jk}$, as dictated by Eq.\eqref{cohrjk}.
\begin{widetext}
\graphicspath{{Plotss//}}
\begin{figure}[htp]
\centering
(a)\includegraphics[width=.3\textwidth]{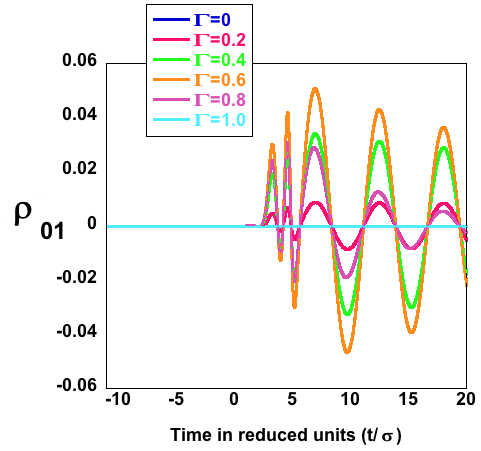}
(b)\includegraphics[width=.3\textwidth]{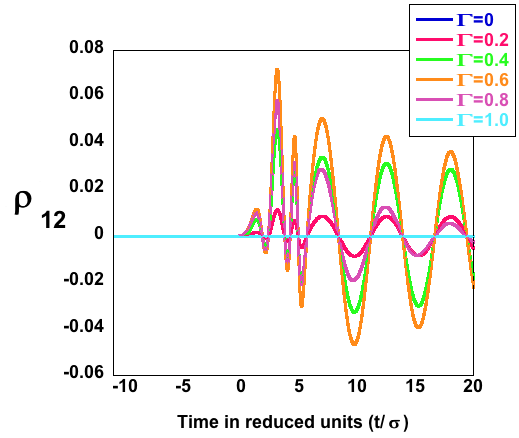}
(c)\includegraphics[width=.3\textwidth]{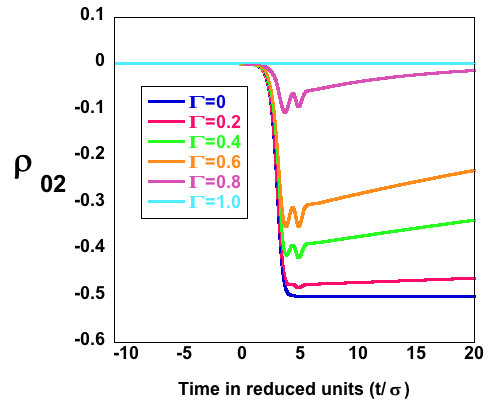}
\label{fig:GammaCohrFSP}
\caption{Plots of the  bare state coherences for different values of $\Gamma$ using FSTIRAP pulse protocol. System is initially prepared to be in the ground state $|0\rangle$  (a) Coherence between the ground state $|0\rangle$ and the intermediate state $|1\rangle$ (b) Coherence between the intermediate state $|1\rangle$ and the target state $|2\rangle$ (c) Coherence between the ground state $|0\rangle$ and the target state $|2\rangle$. Where the $\Gamma=0,0.2,0.4,0.6,0.8,1$ respectively represent for values $\Gamma=0, \frac{1}{500\sigma}, \frac{1}{100\sigma},\frac{1}{50\sigma},\frac{1}{10\sigma},\frac{1}{\sigma}$}
\label{fig:GammaCohrFSP}
\end{figure}
\end{widetext}

Fo the case of FSTIRAP  the maximum transfer efficiency is $\frac{1}{2}$, on to states $|0\rangle$ and the target state $|2\rangle$, which decreases to value of $\frac{1}{3}$ as the noise from environment increases. That means, as can be seen in Figs. \ref{fig:GammaPopFSP}, the population, at the end of the interaction, on states $|0\rangle, |2\rangle$ decreases from $\frac{1}{2}\rightarrow\frac{1}{3}$, consequently this has the implication from Eq.\eqref{cohrjk} that the coherence $\rho_{02}$ would decrease as the noise increase and this is seen in Fig. (\ref{fig:GammaCohrFSP}c).

Interestingly the decrease in population due to noise on states $|0\rangle, |2\rangle$ implies that there is an increase in the population on state $|1\rangle$ from $0\rightarrow\frac{1}{3}$, and consequently this leads the coherences  $\rho_{01}, \rho_{12}$ to increase, albeit small. This is so because there is an increase in population of state $|1\rangle$, at the end of the interaction, while simultaneously decreasing the populations on states $|0\rangle, |2\rangle$ resulting in net increase of the value of coherences up to some optimum value from which it starts to decrease as the system becomes incoherent due to strong environmental noise, this effect is seen in Figs.(\ref{fig:GammaCohrFSP}a) and (\ref{fig:GammaCohrFSP}b).\\

To further show effect of noise to either of the populations and/or coherences in both protocol, STIRAP and FSTIRAP, in Fig.(\ref{fig:GammaRlxCohrFSP}) plots of the final population/coherence versus the coupling strength is given  In line with our previous assertion, it is shown that the populations equilibrates to $\frac{1}{3}$ as the coupling is increased. Moreover it is noted  that the population transferred to the target state $|2\rangle$ decreases exponentially with electron-phonon coupling. For FSTIRAP protocol as the final population of states $|0\rangle$ and $|2\rangle$ decreases with increasing strength of coupling, there is an increase in population in state $|1\rangle$ until the whole processes becomes fully incoherent. It is seen here that the coherences $\rho_{01},\rho_{12}$ increases enough to be experimentally measured (or noticed) until some optimum value of coupling strength, after which it is noticed to decrease. 
\begin{widetext}
\graphicspath{{Plotss//}}
\begin{figure}[htp]
\centering
(a)\includegraphics[width=.3\textwidth]{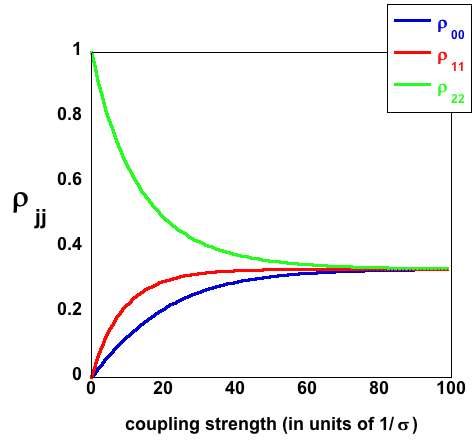}
(b)\includegraphics[width=.3\textwidth]{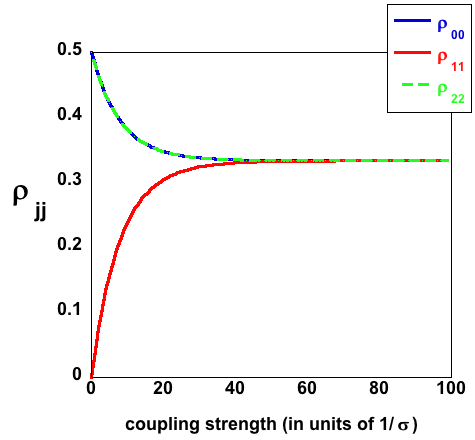}\hfill
(c)\includegraphics[width=.3\textwidth]{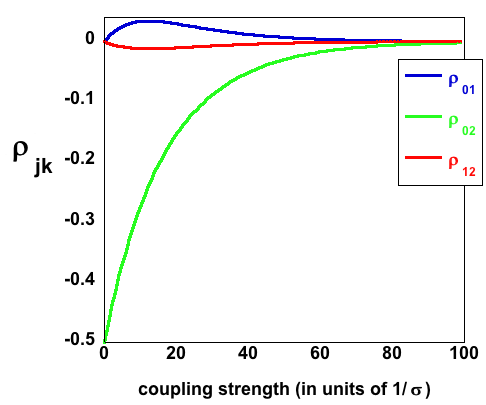}
(d)\includegraphics[width=.3\textwidth]{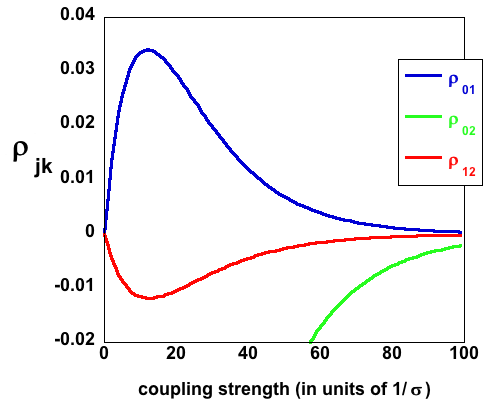}
\caption{Plots of the  final bare state population and coherences versus $\Gamma$ for both STIRAP and FSTIRAP pulse protocol. System is initially prepared to be in the ground state $|0\rangle$  (a) Final population of states $|0\rangle,|1\rangle,|2\rangle$ using the STIRAP protocol  (b) Final population of states $|0\rangle,|1\rangle,|2\rangle$ when the FSTIRAP protocol is used (c) Real parts of the final coherence. (d) zoomed in plots of Real parts of the final coherence. Where $\Gamma=\frac{n}{500\sigma}$ with $n=0,1,2,\dots,100$ represent the strength of the coupling }
\label{fig:GammaRlxCohrFSP}
\end{figure}
\end{widetext}

So far the dynamics and the discussions mainly is focused on a system at zero-temperature with small coupling strength taken into account. In what follows the temperature and the coupling strength will be varied to see a clear picture of what is happening. Because the interest lies in how the coherences are affected by change of either of these parameters, i.e. temperature and/or coupling strength, focus is made on the FSTIRAP pulse profile as it is with such pulse profile that the coherence are created. To this end it is shown below, in Figs.(\ref{fig:LambdaNFSPpop}), the plots for populations on states $|0\rangle,|1\rangle$ as both the temperature and coupling strength changes. While the populations increase (for state $|1\rangle$) and decrease (for states $|0\rangle,|2\rangle$) leading to equilibrium, it is to be noted  from the results that large coupling strength or high temperature brings the system to equilibrium fast. Therefore interesting dynamics is revealed around low values of temperature and small coupling strengths.

For the sake of clarity the populations, of states $|1\rangle,|2\rangle$, are replotted as seen in Figs.(\ref{fig:Pop_vs_LN110}), this is done by fixing one of the parameters while varying the other parameter. consequently the populations are plotted versus the varying parameter. This is to mean that, if for instance the value of the temperature is fixed to be either low or high value, i.e. $N=1, 10$, respectively, plots of population will be versus coupling strength, i.e $(\frac{\lambda}{\omega_{ph}})^2$, and vice versa. In these plots the dashed lines represent populations at higher temperature (or large coupling strength) while the solid lines represent the population for lower temperature (or small coupling strength).
\begin{widetext}
\graphicspath{{Plotsenv//}}
\begin{figure}[htp]
\centering
(a)\includegraphics[width=.4\textwidth]{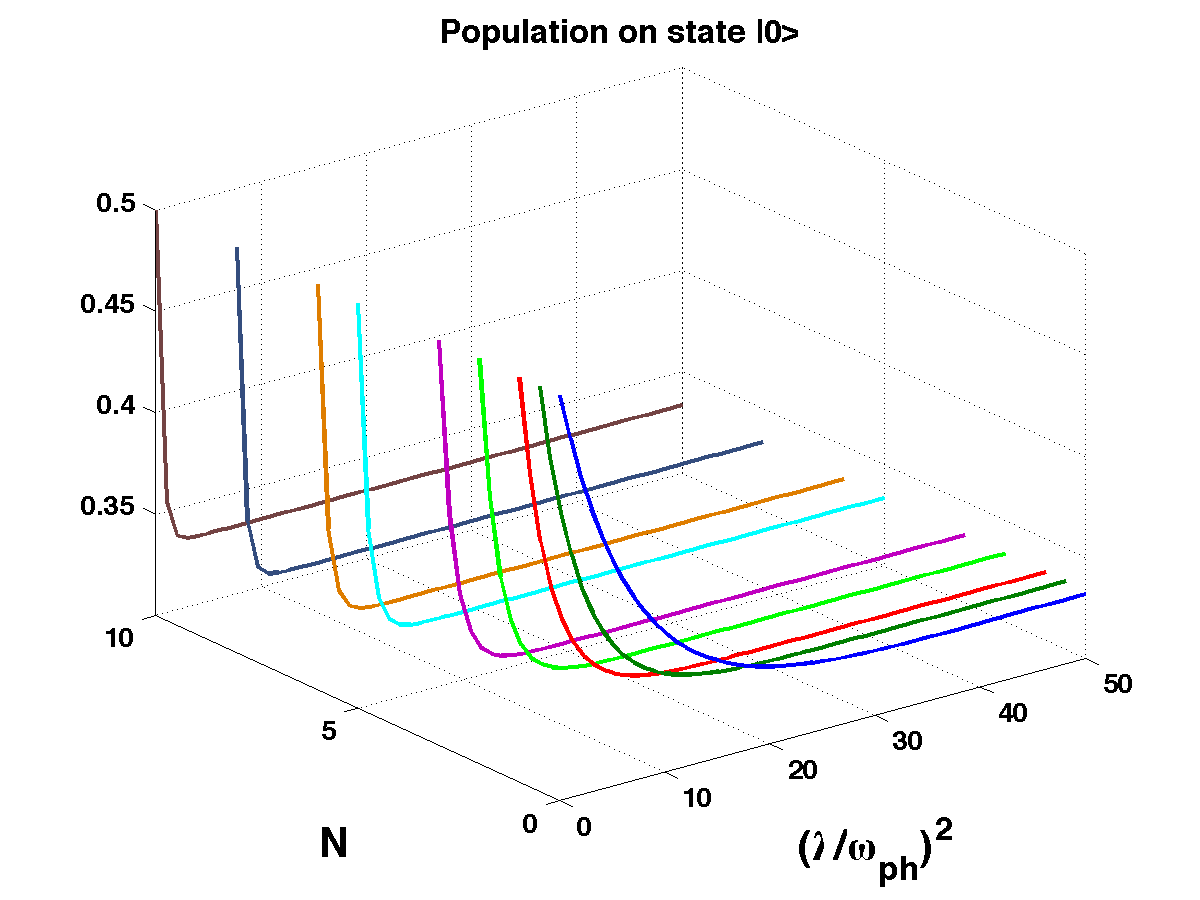}
(b)\includegraphics[width=.5\textwidth]{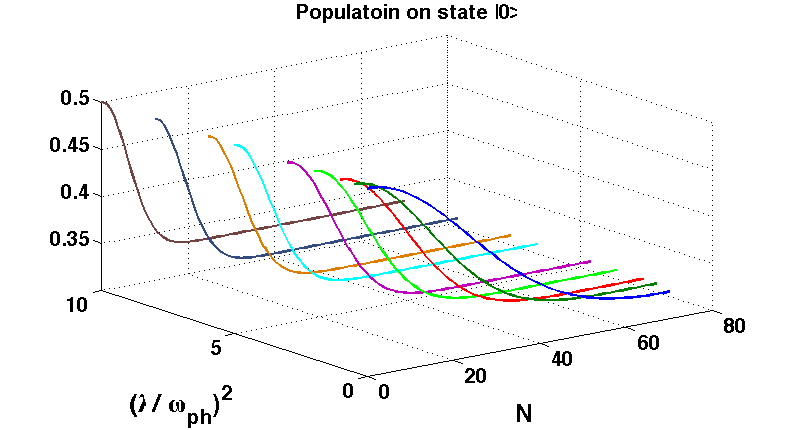}\hfill
(c)\includegraphics[width=.4\textwidth]{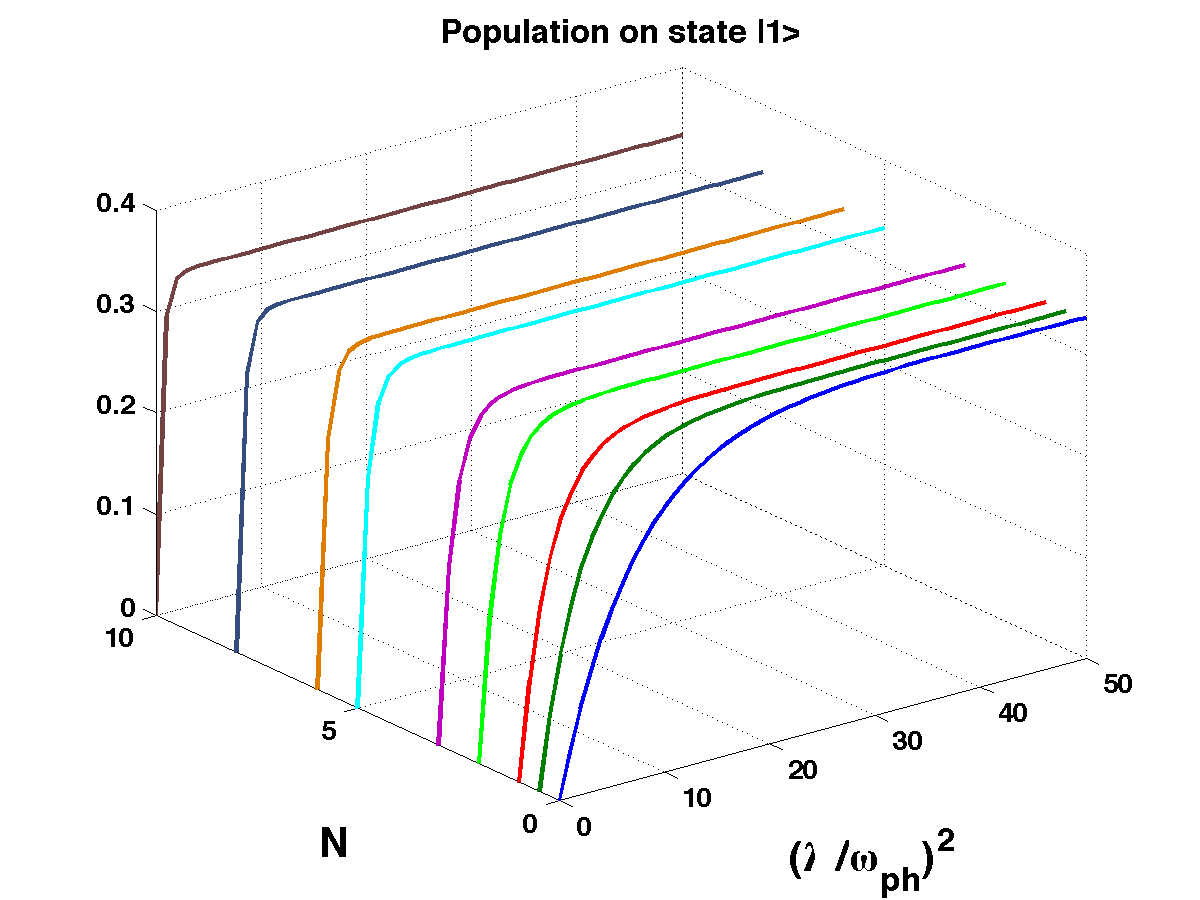}
(d)\includegraphics[width=.5\textwidth]{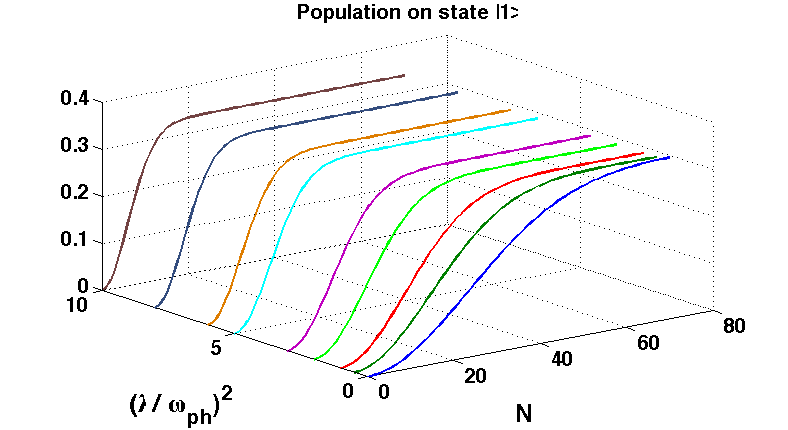}
\caption{Plots of the  final bare state population of states $|0\rangle, |1\rangle$ for FSTIRAP pulse protocol. System is initially prepared to be in the ground state $|0\rangle$  (a) Final population of states $|0\rangle$ vs. coupling strength $(\frac{\lambda}{\omega_{ph}})^2$ at fixed values of temperature  (b) Final population of states $|0\rangle$ vs. temperature for selected values of coupling strength $(\frac{\lambda}{\omega_{ph}})^2$. (c) Final population of states $|1\rangle$ vs. coupling strength $(\frac{\lambda}{\omega_{ph}})^2$ at fixed values of temperature  (d) Final population of states $|1\rangle$ vs. temperature for selected values of coupling strength $(\frac{\lambda}{\omega_{ph}})^2$.Where the colors of the curves represent strength of coupling (and/or values of temperature). Blue stands for small coupling (or low temperature), while brown represent for large coupling (or high temperature).}
\label{fig:LambdaNFSPpop}
\end{figure}
\end{widetext}

Interestingly the decrease (or increase) of the populations versus the parameters is different. It is shown that the change is exponential when population is plotted versus the change with the coupling strength (at a particular temperature); whereas the change is monotonic when it is plotted versus temperature  (for fixed value of coupling strength). Another feature worth stressing here is that at high temperature the system immediately equilibrates and and consequentially no interesting physics is revealed. Following this it is recommended, if one is to see interesting physics, to tune the parameters to low temperature and/or with small coupling strengths. In tuning the parameters to lower regime (low temperature and small coupling strengths), the environment is allowed to stir the system allowing enough time before system equilibrates.  

\begin{widetext}
\graphicspath{{Plotsenv//}}
\begin{figure}[htp]
\centering
(a)\includegraphics[width=.3\textwidth]{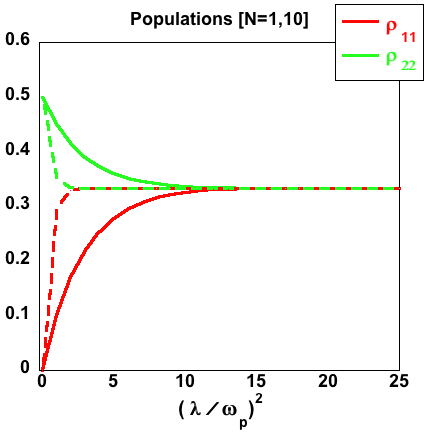}
(b)\includegraphics[width=.3\textwidth]{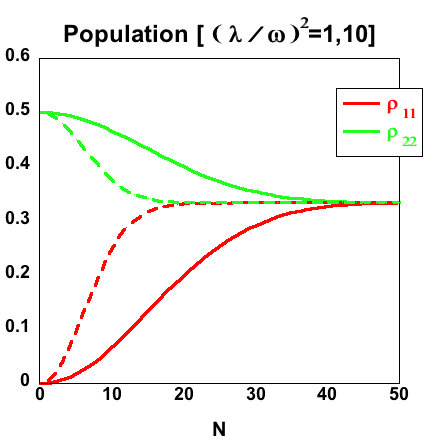}
\caption{Plots of the  final bare state population of states $|1\rangle, |2\rangle$ using FSTIRAP pulse protocol. System is initially prepared to be in the ground state $|0\rangle$  (a) Final population of $\rho_{11}, \rho_{22}$ vs. coupling strength $(\frac{\lambda}{\omega_{ph}})^2$ for low temperature (solid lines) and high temperature (dashed lines)  (b) Final population of $\rho_{11}, \rho_{22}$ vs. temperature for small coupling strength (solid lines) and large coupling strength (dashed lines). Where red represent $\rho_{11}$ and green represent $\rho_{22}$ }
\label{fig:Pop_vs_LN110}
\end{figure}
\end{widetext}

For coherences, similar to the populations, the low temperature and small coupling regime is the interesting region. This can be shown in Figs.(\ref{fig:LambdaNFSPchr}) where it is seen that changes in either the temperature and/or coupling strength first increases the coherences, for instance see $\rho_{01}$ in Figs.(\ref{fig:LambdaNFSPchr}), followed by a decrease. At higher temperature, specifically, these increase and decrease phenomenon is so fast the it is revealed as narrowing in the line shape.  In comparison the line shape is relatively broadened even for large values of coupling strength. This means that the life span of the coherences is seen to be shortened while increasing the temperature or the strength of the coupling. On top of which the magnitude of the coherence also decreases as either of the parameters increase. 

\begin{widetext}
\graphicspath{{Plotsenv//}}
\begin{figure}[htp]
\centering
(a)\includegraphics[width=.4\textwidth]{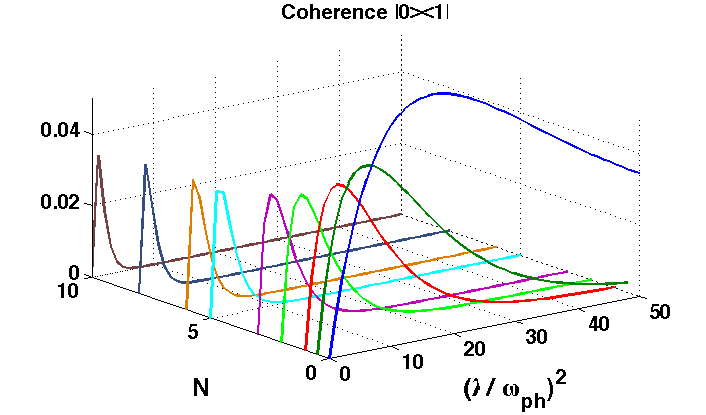}
(b)\includegraphics[width=.4\textwidth]{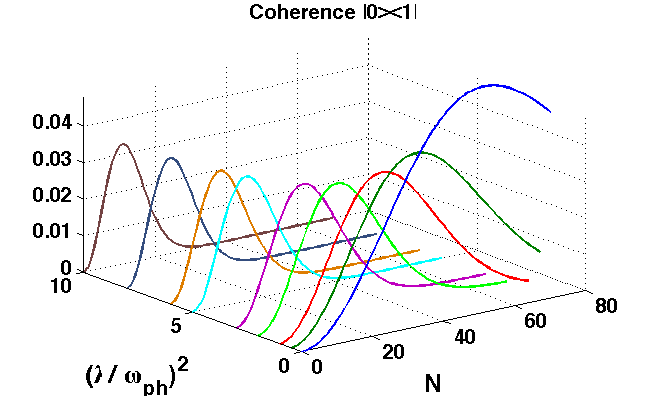}
\caption{Plots of the  coherences between states $|0\rangle, |1\rangle$ using FSTIRAP pulse protocol. System is initially prepared to be in the ground state $|0\rangle$  (a) Real part of coherence $|0\rangle\langle1|$ vs. coupling strength $(\frac{\lambda}{\omega_{ph}})^2$ at fixed values of temperature  (b) Real part of coherence $|0\rangle\langle1|$ vs. temperature for selected values of coupling strength $(\frac{\lambda}{\omega_{ph}})^2$. Where the colors of the curves represent strength of coupling (and/or values of temperature). Blue stands for small coupling (or low temperature), while brown represent for large coupling (or high temperature).}
\label{fig:LambdaNFSPchr}
\end{figure}
\end{widetext}

To conclude it is seen that the effect of temperature on one hand and influence of the coupling strength on the other hand, both of these parameters seen to decrease the magnitude of the coherences as either of the parameters increase. Furthermore the decrease in the population of states $|0\rangle, |2\rangle$ and an increase in population in state $|1\rangle$ is observed to follow an exponential decay with the coupling strength at specific values of temperatures; on the other side these change in population of states versus temperature for fixed values of coupling strength shows monotonic behavior. From this it follows that the life span of the populations (and thereby of the coherences) is greatly affected by the temperature change. This is to say that the system equilibrates pretty fast at higher temperatures in comparison to considering higher coupling strengths.

\section{Conclusion}
\label{sec:conclusion}
This work explored the effect of phonon coupling on the transfer of population and creation of coherence using \emph{fractional} stimulated Raman adiabatic passage (FSTIRAP) which is a variant of stimulated Raman adiabatic passage (STIRAP).  Although the phonon is assumed to be coupled only to the intermediate state, it is shown to be coupled to the other two states by dipolar system-environment interaction, inducing phonon coupling to the other states which are not directly in contact with the phonon.  

Therefore using STIRAP pulse protocol, it is shown that the populations equilibrates to $\frac{1}{3}$, making the processes fully incoherent, as coupling increased. The incoherent processes arises because the coupled phonon vibrates the states until the whole system equilibrates. The shaky nature of the levels in turn creates inability of complete population transfer as would have been the norm for STIRAP. This difficulty can be viewed in two ways, first the vibrating levels are no more able to retain the population transferred to them because the phonon is being shaking the whole system. Second the path to the target state is no more achievable as it is not stable because of the coupling. Moreover it is noted  that the population transferred to the target state $|2\rangle$ decreases exponentially with electron-phonon coupling. 

For FSTIRAP protocol as the final population of states $|0\rangle$ and $|2\rangle$ decreases with increasing strength of coupling, there is an increase in population in state $|1\rangle$ until the whole processes becomes fully incoherent. It is seen here that the coherences $\rho_{01},\rho_{12}$ increases enough to be experimentally measured (or noticed) until some optimum value of coupling strength, after which it is observed to decrease. 
The decrease in population due to noise on states $|0\rangle, |2\rangle$ consequentially results corresponding increase in the population on state $|1\rangle$, and as a result of which the coherences  $\rho_{01}, \rho_{12}$ is seen to increase, albeit small. On top of this as the strength of the noise increases further, the coupling with phonon completely destroys  the phase correlation, and  the coherences deteriorates. This is a clear indication and an interesting observation where moderate noise is seen creating coherences. The optimum value that enables the creation before it starts destroying the coherences is related to the time scale of the leaky level, i.e $|1\rangle$, to which the phonon is directly coupled. 

\appendix
\renewcommand{\thesection}{A.\arabic{section}}
\numberwithin{equation}{section}
\section{Application of Davies-Sphon Formalism}
\label{sec:applyDaviesSphon}
It is worth pointing out that in the rotating framework with a slowly-varying system Hamiltonian and a time dependent system-environment interaction term consequently one can employ the general formalism of Davies and Sphon to derive the master equation. To this end regard the system hamiltonian $H_{sys}$ as constant and the system is interacting with an environment which contains an oscillating term. Because the environmental correlation time is much smaller than the timescales of the Hamiltonian change, therefore it can be treated as if the system Hamiltonian  is constant by assuming that the system Hamiltonian is time-independent during the derivation of the master equation and including the time dependence of the transition operators latter.  Later in the next section we will employ these formalism, i.e Davies-Sphon to the system at hand.
Hence considering a time-independent system Hamiltonian $H_{sys}$ and the following time-dependent system-bath interaction Hamiltonian
 \begin{equation}
\begin{aligned}
V=&\sum_{\alpha}\left(A_{\alpha}^{+}e^{i\omega_{\alpha}t}+A_{\alpha}^{-}e^{-i\omega_{\alpha}t}\right)\otimes B_{\alpha}
\end{aligned}
\label{v}
\end{equation}
Introduce for each Bohr frequency $\omega$
\begin{equation}
\begin{aligned}
A_{\alpha}^{\pm}\left(\omega\right)=&\sum_{\epsilon'-\epsilon=\omega}\Pi\left(\epsilon\right)A_{\alpha}^{\pm}\Pi\left(\epsilon'\right)
\end{aligned}
\label{Aalphapm}
\end{equation}
where $\Pi\left(\epsilon\right)$ is the projector operator on the subspace of the system Hilbert space corresponding to the energy eigenvalue $\epsilon$ and the sum is extended overall the couples of energies $\epsilon$ and $\epsilon'$ such that $\epsilon'-\epsilon=\omega$. The operators defined this way satisfy both of the following ($H_s=H_{sys}$)
\begin{equation}
\begin{aligned}
\left[H_s, A_{\alpha}^{\pm}\left(\omega\right)\right]=&-\omega_{\alpha}A_{\alpha}^{\pm}\left(\omega\right)\\
\left(A_{\alpha}^{\pm}\left(\omega\right)\right)^\dagger=&A_{\alpha}^{\mp}\left(-\omega\right)
\end{aligned}
\label{AalphaHcom}
\end{equation}
from which follows then
\begin{equation}
\begin{aligned}
e^{iH_s t}A_{\alpha}^{\pm}\left(\omega\right)e^{-iH_s t}=&e^{-i\omega t}A_{\alpha}^{\pm}\left(\omega\right)\\
e^{iH_s t}\left(A_{\alpha}^{\pm}\left(\omega\right)\right)^\dagger e^{-iH_s t}=&e^{i\omega t}\left(A_{\alpha}^{\pm}\left(\omega\right)\right)^\dagger
\end{aligned}
\label{INTHam}
\end{equation}
Another important property is that summing over all the Bohr frequencies (both negative and positive) one re obtains the initial operator
\begin{equation}
\begin{aligned}
A_{\alpha}^{\pm}=&\sum_\alpha A_{\alpha}^{\pm}\left(\omega\right)
\end{aligned}
\end{equation}
In the Schr\"{o}dinger picture it takes the form
\begin{equation}
\begin{aligned}
V=&\sum_{\alpha,\omega}\left(A_{\alpha}^{+}\left(\omega\right)e^{i\omega_{\alpha}t}+A_{\alpha}^{-}\left(\omega\right)e^{-i\omega_{\alpha}t}\right)\otimes B_{\alpha}
\end{aligned}
\label{vschro}
\end{equation}
which in the interaction picture w.r.t $H_s+H_B$
\begin{equation}
\begin{aligned}
V=&\sum_{\alpha,\omega}e^{-i\omega t}\left(A_{\alpha}^{+}\left(\omega\right)e^{i\omega_{\alpha}t}+A_{\alpha}^{-}\left(\omega\right)e^{-i\omega_{\alpha}t}\right)\otimes B_{\alpha}\left(t\right)
\end{aligned}
\label{vINT}
\end{equation}
with the Hermitian conjugate $B_{\alpha}\left(t\right)=e^{iH_B t}B_{\alpha}e^{-iH_B t}$ and $\langle B_{\alpha}\left(t\right)\rangle=0$
\begin{equation}
\begin{aligned}
V=&\sum_{\alpha,\omega}e^{i\omega t}\left(\left(A_{\alpha}^{+}\left(\omega\right)\right)^\dagger e^{-i\omega_{\alpha}t}+\left(A_{\alpha}^{-}\left(\omega\right)\right)^\dagger e^{i\omega_{\alpha}t}\right)\otimes B_{\alpha}^\dagger\left(t\right)
\end{aligned}
\label{vINTconj}
\end{equation}
Then the formal solution of the Liouville equation can be written as 
\begin{equation}
\begin{aligned}
\dot\rho=&\int_0^\infty ds~ tr_B\{V\left(t-s\right)\rho\left(t\right)\rho_B V\left(t\right)-V\left(t\right)V\left(t-s\right)\rho\left(t\right)\rho_B\}\\&+h.c
\end{aligned}
\label{rodot}
\end{equation}
using Eqs.\eqref{vINT} and \eqref{vINTconj}, Eq.\eqref{rodot} takes the form
\begin{widetext}
\begin{equation}
\begin{aligned}
\dot\rho=&\int_0^\infty ds~ tr_B\{e^{-i\omega (t-s)}\left(A_{\beta}^{+}\left(\omega\right)e^{i\omega_\beta (t-s)}+A_{\beta}^{-}\left(\omega\right)e^{-i\omega_\beta (t-s)}\right)\rho\left(t\right)\rho_B \\
& \times e^{-i\omega't}\left(\left(A_{\alpha}^{+}\left(\omega'\right)\right)^\dagger e^{-i\omega_\alpha t}+\left(A_{\alpha}^{-}\left(\omega'\right)\right)^\dagger e^{i\omega_\alpha t }\right)B_{\alpha}^\dagger\left(t\right)B_{\beta}\left(t-s\right)\\
&-e^{-i\omega't}\left(\left(A_{\alpha}^{+}\left(\omega'\right)\right)^\dagger e^{-i\omega_\alpha t}+\left(A_{\alpha}^{-}\left(\omega'\right)\right)^\dagger e^{i\omega_\alpha t }\right)\left(A_{\beta}^{+}\left(\omega\right)e^{i\omega_\beta (t-s)}+A_{\beta}^{-}\left(\omega\right)e^{-i\omega_\beta (t-s)}\right)\\
& \times e^{-i\omega (t-s)} \rho\left(t\right)\rho_B B_{\alpha}^\dagger\left(t\right)B_{\beta}\left(t-s\right)\}
\end{aligned}
\label{rhodot}
\end{equation}
\end{widetext}
it then follows that 
\begin{widetext}
\begin{equation}
\begin{aligned}
\dot\rho=&\int_0^\infty ds~\Big(e^{i(\omega'-\omega+\omega_\beta-\omega_\alpha)t}e^{i(\omega-\omega_\beta)s}\left(A_{\beta}^{+}\left(\omega\right)\rho\left(A_{\alpha}^{+}\left(\omega'\right)\right)^\dagger-\left(A_{\alpha}^{+}\left(\omega'\right)\right)^\dagger A_{\beta}^{+}\left(\omega\right)\rho\right)\\
+&e^{i(\omega'-\omega+\omega_\beta+\omega_\alpha)t}e^{i(\omega-\omega_\beta)s}\left(A_{\beta}^{+}\left(\omega\right)\rho\left(A_{\alpha}^{-}\left(\omega'\right)\right)^\dagger-\left(A_{\alpha}^{-}\left(\omega'\right)\right)^\dagger A_{\beta}^{+}\left(\omega\right)\rho\right)\\
+&e^{i(\omega'-\omega-\omega_\beta-\omega_\alpha)t}e^{i(\omega+\omega_\beta)s}\left(A_{\beta}^{-}\left(\omega\right)\rho\left(A_{\alpha}^{+}\left(\omega'\right)\right)^\dagger-\left(A_{\alpha}^{+}\left(\omega'\right)\right)^\dagger A_{\beta}^{-}\left(\omega\right)\rho\right)\\
+&e^{i(\omega'-\omega-\omega_\beta+\omega_\alpha)t}e^{i(\omega+\omega_\beta)s}\left(A_{\beta}^{-}\left(\omega\right)\rho\left(A_{\alpha}^{-}\left(\omega'\right)\right)^\dagger-\left(A_{\alpha}^{-}\left(\omega'\right)\right)^\dagger A_{\beta}^{-}\left(\omega\right)\rho\right)\Big)\\ &\times tr_B\left(B_{\alpha}^\dagger\left(t\right)B_{\beta}\left(t-s\right)\rho_B\right)+h.c
\end{aligned}
\label{rhodot2}
\end{equation}
\end{widetext}
with the following definition
\begin{equation}
\begin{aligned}
\Gamma_{\alpha\beta}^{++}\left(\omega\right)=&\Gamma_{\alpha\beta}^{-+}\left(\omega\right)=&\int_0^\infty ds~e^{i(\omega-\omega_\beta)s}\langle B_{\alpha}^\dagger\left(t\right)B_{\beta}\left(t-s\right)\rangle
\end{aligned}
\label{Gammapp}
\end{equation}
and
\begin{equation}
\begin{aligned}
\Gamma_{\alpha\beta}^{+-}\left(\omega\right)=&\Gamma_{\alpha\beta}^{--}\left(\omega\right)=&\int_0^\infty ds~e^{i(\omega+\omega_\beta)s}\langle B_{\alpha}^\dagger\left(t\right)B_{\beta}\left(t-s\right)\rangle
\end{aligned}
\label{Gammamm}
\end{equation}
it can readily be seen that 
\begin{widetext}
\begin{equation}
\begin{aligned}
\dot\rho=&\sum_{\omega,\omega'}\sum_{\alpha,\beta}e^{i(\omega'-\omega+\omega_\beta-\omega_\alpha)t}\Gamma_{\alpha\beta}^{++}\left(\omega\right)\left(A_{\beta}^{+}\left(\omega\right)\rho\left(A_{\alpha}^{+}\left(\omega'\right)\right)^\dagger-\left(A_{\alpha}^{+}\left(\omega'\right)\right)^\dagger A_{\beta}^{+}\left(\omega\right)\rho\right)\\
+&\sum_{\omega,\omega'}\sum_{\alpha,\beta}e^{i(\omega'-\omega+\omega_\beta+\omega_\alpha)t}\Gamma_{\alpha\beta}^{-+}\left(\omega\right)\left(A_{\beta}^{+}\left(\omega\right)\rho\left(A_{\alpha}^{-}\left(\omega'\right)\right)^\dagger-\left(A_{\alpha}^{-}\left(\omega'\right)\right)^\dagger A_{\beta}^{+}\left(\omega\right)\rho\right)\\
+&\sum_{\omega,\omega'}\sum_{\alpha,\beta}e^{i(\omega'-\omega-\omega_\beta-\omega_\alpha)t}\Gamma_{\alpha\beta}^{+-}\left(\omega\right)\left(A_{\beta}^{-}\left(\omega\right)\rho\left(A_{\alpha}^{+}\left(\omega'\right)\right)^\dagger-\left(A_{\alpha}^{+}\left(\omega'\right)\right)^\dagger A_{\beta}^{-}\left(\omega\right)\rho\right)\\
+&\sum_{\omega,\omega'}\sum_{\alpha,\beta}e^{i(\omega'-\omega-\omega_\beta+\omega_\alpha)t}\Gamma_{\alpha\beta}^{--}\left(\omega\right)\left(A_{\beta}^{-}\left(\omega\right)\rho\left(A_{\alpha}^{-}\left(\omega'\right)\right)^\dagger-\left(A_{\alpha}^{-}\left(\omega'\right)\right)^\dagger A_{\beta}^{-}\left(\omega\right)\rho\right)+h.c
\end{aligned}
\label{rhodot3}
\end{equation}
\end{widetext}
This is the most general form of the Born-Markov master equation before Rotating Wave Approximation (RWA) is performed. Under the hypothesis that $\omega_\alpha, \omega_\beta>>\omega, \omega'$, one can single out very clear condition for RWA. The only term which survives are those for which $\omega_\alpha$ and $\omega_\beta$ appear in the combination $\omega_\alpha-\omega_\beta$, with $\alpha=\beta$ and $\omega=\omega'$
\begin{widetext}
\begin{equation}
\begin{aligned}
\dot\rho=&\sum_{\omega}\sum_{\alpha}\Gamma_{\alpha\alpha}^{++}\left(\omega\right)\left(A_{\alpha}^{+}\left(\omega\right)\rho\left(A_{\alpha}^{+}\left(\omega\right)\right)^\dagger-\left(A_{\alpha}^{+}\left(\omega\right)\right)^\dagger A_{\alpha}^{+}\left(\omega\right)\rho\right)\\
+&\sum_{\omega}\sum_{\alpha}\Gamma_{\alpha\alpha}^{--}\left(\omega\right)\left(A_{\alpha}^{-}\left(\omega\right)\rho\left(A_{\alpha}^{-}\left(\omega\right)\right)^\dagger-\left(A_{\alpha}^{-}\left(\omega\right)\right)^\dagger A_{\alpha}^{-}\left(\omega\right)\rho\right)+h.c
\end{aligned}
\label{rhodot4}
\end{equation}
\end{widetext}
Now with one sided Fourier transforms
\begin{equation}
\begin{aligned}
\Gamma_{\alpha\beta}\left(\omega\right)=&\int_0^\infty ds~e^{i\omega s}\langle B_{\alpha}^\dagger\left(t\right)B_{\beta}\left(t-s\right)\rangle
\end{aligned}
\end{equation}
of the reservoir correlation function
\begin{equation}
\begin{aligned}
\langle B_{\alpha}^\dagger\left(t\right)B_{\beta}\left(t-s\right)\rangle\equiv tr_B\{B_{\alpha}^\dagger\left(t\right)B_{\beta}\left(t-s\right)\rho_B\}
\end{aligned}
\end{equation}
showing that the quantities $\Gamma_{\alpha\beta}\left(\omega\right)$ do not depend on time. It is convenient to decompose the Fourier transforms of the reservoir correlation function as follows
\begin{equation}
\begin{aligned}
\Gamma_{\alpha\beta}\left(\omega\right)=&\frac{1}{2}\gamma_{\alpha\beta}\left(\omega\right)+iS_{\alpha\beta}\left(\omega\right)
\end{aligned}
\end{equation}
where for fixed $\omega$ the coefficients 
\begin{equation}
\begin{aligned}
S_{\alpha\beta}\left(\omega\right)=&\frac{1}{2i}\left(\Gamma_{\alpha\beta}\left(\omega\right)-\Gamma_{\beta\alpha}^*\left(\omega\right)\right)
\end{aligned}
\end{equation}
for Hermitian matrix and the matrix is defined by 
\begin{equation}
\begin{aligned}
\gamma_{\alpha\beta}=&\Gamma_{\alpha\beta}\left(\omega\right)+\Gamma_{\beta\alpha}^*\left(\omega\right)=\int_{-\infty}^\infty ds~e^{i\omega s}\langle B_{\alpha}^\dagger\left(s\right)B_{\beta}\left(0\right)\rangle
\end{aligned}
\end{equation}
is positive. Thus Eq.\eqref{rhodot4} can be rewritten as, coming back to Schr\"{o}dinger picture
\begin{widetext}
\begin{equation}
\begin{aligned}
\dot\rho=&-i\left[H_s,\rho\right]+\sum_{\omega}\sum_{\alpha}\gamma_{\alpha\alpha}^{++}\left(\omega\right)\left(A_{\alpha}^{+}\left(\omega\right)\rho\left(A_{\alpha}^{+}\left(\omega\right)\right)^\dagger-\frac{1}{2}\{\left(A_{\alpha}^{+}\left(\omega\right)\right)^\dagger A_{\alpha}^{+}\left(\omega\right),\rho\}\right)\\
+&\sum_{\omega}\sum_{\alpha}\gamma_{\alpha\alpha}^{--}\left(\omega\right)\left(A_{\alpha}^{-}\left(\omega\right)\rho\left(A_{\alpha}^{-}\left(\omega\right)\right)^\dagger-\frac{1}{2}\{\left(A_\alpha^{-}\left(\omega\right)\right)^\dagger A_{\alpha}^{-}\left(\omega\right),\rho\}\right)+h.c
\end{aligned}
\label{rhodot5}
\end{equation}
\end{widetext}
where
\begin{equation}
\begin{aligned}
\gamma_{\alpha\alpha}^{++}\left(\omega\right)=&2\Re\left(\Gamma_{\alpha\alpha}^{++}\left(\omega\right)\right)\\
\gamma_{\alpha\alpha}^{--}\left(\omega\right)=&2\Re\left(\Gamma_{\alpha\alpha}^{--}\left(\omega\right)\right)
\end{aligned}
\end{equation}
since in rotating frame there is a slowly varying system Hamiltonian and time-dependent system-environment interaction term one can use the preceding formalism to derive the master equation that describes the dynamics of the system at hand.

\section{Adiabatic Hamiltonian}
\label{sec:Hadiabatic}
We here want to find the form of the system Hamiltonian $H_s$. To this end it will be insightful to work  in the adiabatic basis. When the two-photon resonance condition $\hbar\left(\Delta_p-\Delta_s\right)=0$ is fulfilled, it is already noted the system Hamiltonian to be of the form 
\begin{equation}
\begin{aligned}
\hat H_{sys}\left(t\right)=&\frac{\hbar}{2} \begin{pmatrix}
0 & \Omega_P\left(t\right) & 0 \\
\Omega_P\left(t\right) & 2\Delta & \Omega_S\left(t\right) \\
0 & \Omega_S\left(t\right)  & 0 \end{pmatrix}
\label{Hamil}
\end{aligned}
\end{equation} 
It is worth recalling that for the three level system under consideration, the transition between levels $|0\rangle$ and $|2\rangle$ is dipole forbidden, level $|1\rangle$ can be off-resonance by certain detuning.  Introducing a unitary transformation $|\Psi\rangle=U|\Phi\left(t\right)\rangle$ \cite{Vitanov:1999aa}, by making the expansion of  the wave function in terms of  the diabatic states $\{|0\rangle$, $|1\rangle$, $\& |2\rangle\}$ (also known as the bare states) in terms of the time dependent probability amplitude,  written in vector form as $C_j\left(t\right)=\big(C_0,C_1,C_2 \big)^T$ :
\begin{equation}
\begin{aligned}
|\Psi\rangle=&\sum_{j=0}^2  C_j \left(t\right) |j\rangle 
\end{aligned}
\end{equation}
The unitary matrix $U$ diagonalises the Hamiltonian Eq.\eqref{Hamil} into $D=U^\dagger H U$, adiabatic Hamiltonian, where the transformation matrix $U$ is given by \cite{Vitanov:1999aa}
\begin{equation}
\begin{aligned}
U=& \begin{pmatrix}
\sin\phi\sin\theta & \cos\phi & \sin\phi\cos\theta \\
\cos\theta & 0 & -\sin\theta\\
\sin\theta\cos\phi & -\sin\phi  & \cos\phi\cos\theta \end{pmatrix}
\label{U}
\end{aligned}
\end{equation} 
After diagonalising the Hamiltonian Eq.\eqref{Hamil} it is readily obtained that the new basis vectors $\{|a_+\rangle,|a_0\rangle,|a_-\rangle\}$  corresponds to the eigenvalues $\{\omega_+,\omega_0,\omega_-\}$ respectively where
\begin{subequations}
\begin{align}
\omega_+=&\frac{1}{2}\big(\Delta+\sqrt{\Delta^2+\Omega^2}\big)=\Omega\cot\phi,\\
\omega_0=&0, \\
\omega_-=&\frac{1}{2}\big(\Delta-\sqrt{\Delta^2+\Omega^2}\big)=-\Omega\tan\phi
\label{evalue}
\end{align}
\end{subequations} 
with $\Omega^2=\Omega_P^2+\Omega_S^2$
where we recall the mixing angles are defined to be 
\begin{subequations}
\begin{align}
\tan\theta=&\frac{\Omega_P\left(t\right)}{\Omega_S\left(t\right)},\label{angle1}\\
\tan 2\phi=&\frac{\Omega\left(t\right)}{\Delta}\label{angle2}
\end{align}
\end{subequations} 
If we denote the wave function for  adiabatic states by $|\Phi\rangle=\{|a_+\rangle,|a_0\rangle,|a_-\rangle\}$ and for the diabatic states by $|\Psi\rangle=\{|0\rangle,|1\rangle,|2\rangle\}$, we see that with the help of the transformation $|\Phi\rangle=U|\Psi\rangle$  the Schr\"{o}dinger equation of the state vector given by 
\begin{equation}
\begin{aligned}
\imath\hbar \frac{d }{dt}|\Phi\rangle=&\hat H_{Non-adiabatic}|\Phi\rangle,
\end{aligned}
\end{equation}
takes the following form
\begin{equation}
\begin{aligned}
\imath\hbar \frac{d |\Psi\rangle}{dt}=&\Big(\underbrace{U^\dagger\hat H U-\imath\hbar U^\dagger\frac{d U}{dt}}_{H_{Non-adiabatic}}\Big)|\Psi\rangle\\
\end{aligned}
\end{equation}
where the non-adiabatic Hamiltonian is given by (We note in the transformed notation $H_{Non-adiabatic}=H_s$ or $H_{Adiabatic}=H_s$ where $H_s$ is of Eq.\eqref{Hamil}) 
\begin{equation}
\begin{aligned}
H_{Non-adiabatic}=& \begin{pmatrix}
\Omega\cot\phi & \imath\dot\theta\sin\phi & \imath\dot\phi \\
-\imath\dot\theta\sin\phi & 0 & -\imath\dot\theta\cos\phi  \\
-\imath\dot\phi & \imath\dot\theta\cos\phi    &- \Omega\tan\phi \end{pmatrix}
\label{HnonA}
\end{aligned}
\end{equation} 
where the over-dot means time derivative. Differentiating Eqs.\eqref{angle1}  and \eqref{angle2} with respect to time and using chain rule of differentiation yields
\begin{subequations}
\begin{align}
\dot\theta=&\frac{\dot\Omega_P\Omega_S-\dot\Omega_S\Omega_P}{\Omega_P^2+\Omega_S^2} \label{tetadot}\\
\dot\phi=&\frac{1}{2}\Big[\frac{\Delta}{\Omega}\Big(\frac{\dot\Omega_P\Omega_P+\dot\Omega_S\Omega_S}{\Delta^2+\Omega^2}\Big)\Big] \label{fidot}
\end{align}
\end{subequations}
The off diagonal terms in Eq.\eqref{HnonA}  tells us that there is coupling between the adiabatic states, i.e between $|a_j\rangle $ and $|a_k\rangle $ where $j,k=+,0,-$ and $j\neq k$, this term arose from $\imath\hbar U^\dagger\frac{d U}{dt}$ . For an adiabatic system, the time derivative of the evolution matrix $U$, is zero, if so we would note that off diagonal elements now are zero and our Hamiltonian would reduce to be
\begin{equation}
\begin{aligned}
H_{Adiabatic}=& \begin{pmatrix}
\Omega\cot\phi & 0 & 0 \\
0 & 0 & 0  \\
0 & 0  &  -\Omega\tan\phi \end{pmatrix}=& \begin{pmatrix}
\omega_+ & 0 & 0 \\
0 & \omega_0 & 0  \\
0 & 0  & \omega_- \end{pmatrix}
\end{aligned}
\end{equation} 
Quantum mechanics tells us that adiabatic evolution of the states is assured when the rate of non-adiabatic coupling is small compared to the separation of the corresponding eigenvalues, i.e., when \cite{shore98}
\begin{equation}
\begin{aligned}
\Large |\Big\langle a_\pm \Large |\frac{d}{dt} \Large|a_0\Big\rangle\Large|^2\ll&|\omega_0-\omega_\pm |^2
\label{adiabc}
\end{aligned}
\end{equation}
Insufficient coupling by the coherent radiation fields may prevent the state vector $|\Psi\rangle$ from adiabatically following the evolution of the trapped state $|a_0\rangle$, and loss of population due to non-adiabatic transfer   $|a_+\rangle$ and  $|a_-\rangle$ may occur. The condition for adiabatic following can be derived from Quantum Mechanics 
\begin{equation}
\begin{aligned}
\frac{\Large |\Big\langle m \Large |\frac{d}{dt} \Large|n\Big\rangle\Large|}{\big(E_m-E_n\big)^2}\ll&1    & \forall m,n
\end{aligned}
\end{equation}
where $E_k$ is the $k^{th}$ instantaneous eigenvalues of $H$. 

The Hamiltonian matrix element fro non-adiabatic coupling between state $|a_0\rangle$ and either one of $|a_+\rangle$ or  $|a_-\rangle$ is given by $\langle a_\pm|\dot a_0\rangle$.
non-adiabatic coupling is small if this matrix element is small compared to the field induced splitting $|\omega_\pm-\omega_0 |$ of the energies of these states, that is 
\begin{equation}
\begin{aligned}
|\langle a_\pm|\dot a_0\rangle|\ll&|\omega_\pm-\omega_0 |
\end{aligned}
\end{equation}

Using Eq.\eqref{U} and the transformation $|\Phi\rangle=U|\Psi\rangle$, we already obtained that
\begin{subequations}
\begin{align}
|a_+\rangle=&\sin\theta\sin\phi |0\rangle+\cos\phi  |1\rangle+ \cos\theta\sin\phi |2\rangle \label{estate1}\\
|a_0\rangle=&\cos\theta |0\rangle-\sin\theta  |2\rangle \label{estate2}\\
|a_-\rangle=&\sin\theta\cos\phi |0\rangle-\sin\phi  |1\rangle+ \cos\theta\cos\phi |2\rangle \label{estate3}
\end{align}
\end{subequations}
Notice that the adiabatic state $|a_0\rangle$ is independent of state $|1\rangle$, which is the leaky state. Therefore following the adiabatic state $|a_0\rangle$ we can achieve the complete population transfer  from the ground state $|0\rangle$ to the target state $|2\rangle$. To get the adiabatic following condition, from Eqs.\eqref{estate1}, \eqref{estate2}, and \eqref{estate3} we have that
 \begin{equation}
\begin{aligned}
|\dot a_0\rangle=&-\dot\theta\sin\theta |0\rangle-\dot\theta\cos\theta  |2\rangle\\
\end{aligned}
\end{equation}
Thus from the adiabatic condition, Eq.\eqref{adiabc}, follows
\begin{equation}
\begin{aligned}
|\langle a_+|\dot a_0\rangle|=&-\dot\theta\sin\phi,   && &
|\langle a_-|\dot a_0\rangle|=&-\dot\theta\cos\phi
\end{aligned}
\end{equation}
From which follows
\begin{equation}
\begin{aligned}
|\dot\theta\frac{\sin^2\phi}{\cos\phi}|\ll&|\Omega|,   && &
|\dot\theta\frac{\cos^2\phi}{\sin\phi}|\ll&|\Omega|
\end{aligned}
\end{equation}
\section{Transition Operators}
\label{sec:jumpingOperators}
We already defined the transition operators to be (see Eq.\eqref{Aalphapm})
\begin{equation}
\begin{aligned}
A_{\alpha}^{\pm}\left(\omega\right)=&\sum_{\epsilon'-\epsilon=\omega}\Pi\left(\epsilon\right)A_{\alpha}^{\pm}\Pi\left(\epsilon'\right)
\end{aligned}
\label{Aalphapma}
\end{equation}
 for our three-level system we note that
\begin{equation}
\begin{aligned}
A_a^+=&|0\rangle\langle 1|, && A_a^-=|1\rangle\langle 0|\\
A_b^+=&|2\rangle\langle 1|, && A_b^-=|1\rangle\langle 2|\\
B_a=&\frac{\lambda^{(a)}}{\omega_{ph}}\left(\hat a^\dagger-\hat a\right),  && B_b=\frac{\lambda^{(b)}}{\omega_{ph}}\left(\hat a^\dagger-\hat a\right)
\end{aligned}
\label{Aalphapmsa}
\end{equation}
where $a, b$ represents transitions $0\leftrightarrow 1,1\leftrightarrow 2$ respectively, and we point out  that for our case $\lambda^{(a)}=\lambda^{(b)}=\lambda$. Furthermore we have obtained the eigenstates of the system Hamiltonian to be (see Eqs.\eqref{estate1},\eqref{estate2}, and \eqref{estate3})
\begin{subequations}
\begin{align}
|a_+\rangle=&\sin\theta\sin\phi |0\rangle+\cos\phi  |1\rangle+ \cos\theta\sin\phi |2\rangle \label{estate1a}\\
|a_0\rangle=&\cos\theta |0\rangle-\sin\theta  |2\rangle \label{estate2a}\\
|a_-\rangle=&\sin\theta\cos\phi |0\rangle-\sin\phi  |1\rangle+ \cos\theta\cos\phi |2\rangle \label{estate3a}
\end{align}
\end{subequations}
Below using Eq.\eqref{Aalphapma} we will obtain the jumping operators, we use the notation $\omega_{km}=\omega_k-\omega_m$, where~$ k,m=+,0,-$
\begin{enumerate}
\item $A_{a}^{+}\left(\omega_{+0}\right)$
\begin{equation}
\begin{aligned}
A_{a}^{+}\left(\omega_{+0}\right)=&\sum_{\omega_{+0}}\Pi\left(\omega_0\right)A_{a}^{+}\Pi\left(\omega_+\right)\\
=&|a_0\rangle\langle a_0|A_{a}^{+}|a_+\rangle\langle a_+|\\
=&\cos\theta\cos\phi|a_0\rangle\langle a_+|
\end{aligned}
\label {Aa+0+}
\end{equation}
\item $A_{a}^{-}\left(\omega_{+0}\right)$
\begin{equation}
\begin{aligned}
A_{a}^{-}\left(\omega_{+0}\right)=&\sum_{\omega_{+0}}\Pi\left(\omega_0\right)A_{a}^{-}\Pi\left(\omega_+\right)\\
=&|a_0\rangle\langle a_0|A_{a}^{-}|a_+\rangle\langle a_+|\\
=&0
\end{aligned}
\label {Aa-01}
\end{equation}
\item following same footsteps  one readily arrives at 
\begin{widetext}
\begin{equation}
\begin{aligned}
A_{b}^{+}\left(\omega_{+0}\right)=&-\sin\theta\cos\phi|a_0\rangle\langle a_+| && A_{b}^{-}\left(\omega_{+0}\right)=0\\
A_{a}^{+}\left(\omega_{0-}\right)=&0 && A_{a}^{-}\left(\omega_{0-}\right)=-\sin\phi\cos\theta|a_-\rangle\langle a_0|\\
A_{a}^{-}\left(\omega_{0+}\right)=&\cos\phi\cos\theta|a_+\rangle\langle a_0| && A_{b}^{+}\left(\omega_{0-}\right)=0\\
A_{b}^{-}\left(\omega_{0+}\right)=&-\cos\phi\sin\theta|a_+\rangle\langle a_0| && A_{b}^{-}\left(\omega_{0-}\right)=\sin\phi\sin\theta|a_-\rangle\langle a_0|\\
A_{a}^{+}\left(\omega_{-0}\right)=&-\sin\phi\cos\theta|a_0\rangle\langle a_-| && A_{a}^{+}\left(\omega_{+-}\right)=\cos^2\phi\sin\theta|a_-\rangle\langle a_+|\\
A_{b}^{+}\left(\omega_{-0}\right)=&\sin\phi\sin\theta|a_0\rangle\langle a_-| && A_{a}^{-}\left(\omega_{+-}\right)=-\sin^2\phi\sin\theta|a_-\rangle\langle a_+|\\
A_{a}^{-}\left(\omega_{-+}\right)=&\cos^2\phi\sin\theta|a_+\rangle\langle a_-| && A_{b}^{+}\left(\omega_{+-}\right)=\cos^2\phi\cos\theta|a_-\rangle\langle a_+|\\
A_{a}^{+}\left(\omega_{-+}\right)=&-\sin^2\phi\sin\theta|a_+\rangle\langle a_-| && A_{b}^{-}\left(\omega_{+-}\right)=-\sin^2\phi\cos\theta|a_-\rangle\langle a_+|\\
A_{b}^{-}\left(\omega_{-+}\right)=&\cos^2\phi\cos\theta|a_+\rangle\langle a_-| \\ 
A_{a}^{+}\left(0\right)=&\sin\phi\cos\phi\sin\theta\left(|a_+\rangle\langle a_+|-|a_-\rangle\langle a_-|\right)\\
A_{b}^{+}\left(\omega_{-+}\right)=&-\sin^2\phi\cos\theta|a_+\rangle\langle a_-| \\ 
A_{b}^{+}\left(0\right)=&\sin\phi\cos\phi\sin\theta\left(|a_+\rangle\langle a_+|-|a_-\rangle\langle a_-|\right)\\
A_{a}^{-}\left(0\right)=&\sin\phi\cos\phi\sin\theta\left(|a_+\rangle\langle a_+|-|a_-\rangle\langle a_-|\right) \\
 A_{b}^{-}\left(0\right)=&\sin\phi\cos\phi\sin\theta\left(|a_+\rangle\langle a_+|-|a_-\rangle\langle a_-|\right)\\
\end{aligned}
\label {Aabconj}
\end{equation}
\end{widetext}
\end{enumerate}
\bibliographystyle{plain}
\bibliography{References}

\end{document}